\documentclass[usenatbib]{mnras}
\usepackage{amsmath,amssymb}
\usepackage[pdftex]{graphicx}

\def\vb#1{\mbox{\boldmath $#1$}}

\newcommand{\percc}{{\rm cm^{-3}}}
\newcommand{\um}{{\rm \mu m}}
\newcommand{\E}[1]{\times 10^{#1}}

\newcommand{\nH}{n_{{\rm H}}}

\newcommand{\nHmax}{n_{\rm H,max}}

\newcommand{\ljeans}{\lambda _{\rm J}}

\newcommand{\MPopIII}{M_{\rm Pop III}}

\newcommand{\tff}{t_{\rm ff}}
\newcommand{\tcol}{t_{\rm col}}

\newcommand{\cs}{c_{\rm s}}
\newcommand{\Mturb}{{\cal M}_{\rm turb}}
\newcommand{\fkep}{f_{\rm Kep}}

\newcommand{\abH}[1]{{\rm [{#1}/H]}}

\newcommand{\Forsterite}{{\rm Mg_2SiO_4}}

\newcommand{\namb}{n_{{\rm amb}}}

\newcommand{\XH}{X_{{\rm H}}}
\newcommand{\Zsun}{{\rm Z_{\bigodot}}}
\newcommand{\Msun}{{\rm M_{\bigodot}}}
\newcommand{\Msunyr}{\Msun/{\rm yr}}

\newcommand{\gammaad}{\gamma _{\rm ad}}
\newcommand{\gammaeff}{\gamma _{\rm eff}}

\defcitealias{Chiaki15}{C15}
\defcitealias{Chiaki16}{C16}
\defcitealias{Chiaki18}{C18}
\defcitealias{Smith15}{S15}
\defcitealias{Hirano14}{H14}
\defcitealias{Ishigaki14}{I14}
\defcitealias{Tominaga14}{T14}
\defcitealias{Marassi15}{M15}

\usepackage{color}
\definecolor{rev}{rgb}{0.8,0.0,0.0}
\definecolor{grn}{rgb}{0.0,0.8,0.0}

\title[Low-metallicity disc fragmentation]
      {Disc fragmentation and oligarchic growth of protostellar systems in low-metallicity gas clouds}

\author[G. Chiaki et al.]
{Gen Chiaki$^{1, 2}$\thanks{E-mail: gen.chiaki@astr.tohoku.ac.jp} and
Naoki Yoshida$^{3,4,5}$
\\
$^{1}$Center for Relativistic Astrophysics, School of Physics, Georgia Institute of Technology, Atlanta, GA 30332, USA \\
$^{2}$ Astronomical Institute, Graduate School of Science, Tohoku University, Aoba, Sendai 980-8578, Japan \\
$^{3}$Department of Physics, School of Science, The University of Tokyo, 7-3-1 Hongo, Bunkyo, Tokyo 113-0033, Japan \\
$^{4}$Kavli Institute for the Physics and Mathematics of the Universe (WPI), UT Institute for Advanced Study, The University of Tokyo, \\
Kashiwa, Chiba 277-8583, Japan \\
$^{5}$Research Center for the Early Universe (RESCEU), School of Science, The University of Tokyo, 7-3-1 Hongo, Bunkyo, Tokyo \\
113-0033, Japan}

\begin{document}

\date{}

\pagerange{\pageref{firstpage}--\pageref{lastpage}} \pubyear{2020}

\maketitle

\label{firstpage}

\begin{abstract}
  We study low-metallicity star formation with a set of high-resolution hydrodynamics
  simulations for
  various gas metallicities over a wide range $0$--$10^{-3} \ \Zsun$.
  Our simulations follow non-equilibrium chemistry and radiative cooling
  by adopting realistic elemental abundance and 
  dust size distribution.
  We examine the condition for the fragmentation of collapsing clouds
  (cloud fragmentation; CF) 
  and of accretion discs (disc fragmentation; DF).
  We find that CF is suppressed due to rapid gas 
  heating accompanied with molecular hydrogen formation even with efficient
  dust cooling for metallicities $\gtrsim 10^{-5} \ \Zsun$.
  Instead, DF occurs 
  in almost all runs regardless of metallicity.
  We also find that, in the accretion discs, the growth of the protostellar systems is overall oligarchic.
  The primary protostar grows
  through the accretion of gas, and secondary protostars 
  form through the interaction of spiral arms or
  the break-up of a rapidly rotating protostar.
  Despite vigorous fragmentation, a large fraction of secondary protostars are destroyed
  through mergers or tidal disruption events.
  For a few hundred years after the first adiabatic core formation,
  only several protostars survive in the disc, and
  the total mass of protostars is $0.52$--$3.8 \ \Msun$.
\end{abstract}



\section{INTRODUCTION}
The standard cosmological model (cold dark matter model) predicts that the
cosmic structure forms hierarchically from small to large systems \citep[e.g.][]{Audouze95}.
According to the model, the first generation of metal-free (Population III; Pop III) stars
form in low-mass dark matter
halos (minihalos) with $10^5$--$10^6 \ \Msun$ at redshift $z\sim 10$--$30$
(Yoshida et al. 2003).
The second generation of metal-poor (Population II; Pop II) stars form
from interstellar gas that is chemically enriched by the first supernovae (SNe).
Pop III/II stars play an important role in the early
phase of galaxy formation.
Massive stars can emit ultraviolet (UV) photons and
suppress star formation, whereas their SN explosions can enrich the interstellar medium (ISM)
with metals/grains that can enhance gas cooling and star formation.
The stellar mass is crucial to determine
the rate of UV emission and the total mass of synthesized heavy elements,
but the characteristic mass of Pop III and Pop II stars is poorly known.

Observations of long-lived stars give constraints
on the lower bound of the initial mass functions (IMFs) of Pop III/II stars
because only stars with masses smaller than $0.8 \ \Msun$ can survive over 10 Gyr until the present day.
So far, no stars with carbon abundances $A({\rm C}) < 6$ or iron abundances ${\rm [Fe/H]} < -5$
have been observed \citep{Yoon16, Yoon18, Placco18},\footnote{The abundance ratio of an 
element A to B is conventionally written as
\[
{\rm [A/B]} = 
\left( A({\rm A}) - A({\rm B}) \right) -
\left( A_{\bigodot} ({\rm A}) - A_{\bigodot} ({\rm B}) \right)
\]
where 
\[
A({\rm A}) = 12 + \log \left( y_{\rm A} \right)
\]
is the absolute abundance of A, and $y_{\rm A}$ is the number
abundance of A relative to hydrogen nuclei.
Throughout this paper, we use the solar abundance $A_{\bigodot} ({\rm A})$ of
\citet{Asplund09}.}
which may suggest that the transition of the typical stellar mass occurs from
massive Pop III to Pop II stars.

Motivated by the recent observations,
researchers have proposed theoretical models of low-mass, low-metallicity 
star formation.
The fragmentation of gas is a key physical process to
reduce the mass scale of finally forming stars.
\citet[][hereafter \citetalias{Chiaki16}]{Chiaki16} found that there are
two distinctive modes of fragmentation: the fragmentation of collapsing clouds
(cloud fragmentation; CF) and of accretion discs (disc fragmentation; DF), and that
these modes occur in two different phases of star formation.

In the first phase, called the {\it collapsing phase}, 
an interstellar gas cloud contracts
in a run-away manner through its self-gravity.
Linear analyses show that, for the gas following an equation of state (EOS)
$p \propto \rho ^{\gammaeff}$ with a polytropic index $\gammaeff$, 
perturbations on a spherically collapsing cloud grow unstably, and filamentary structures develop
for $\gammaeff < 1.097$ \citep{Hanawa00, Lai00}.
When density fluctuations on the filaments grow sufficiently, multiple clumps appear
on the filamentary cloud,  leading CF \citep{Tsuribe06}.
It has been suggested that gas cooling through dust thermal emission is the important 
cooling process even with very low metallicities $\sim 10^{-5} \ \Zsun$
\citep{Omukai00, Schneider03, Omukai05, Dopcke11, SafranekShrader14b, Smith15}.

After the gas in the cloud center becomes optically thick, 
it can no longer collapse because of inefficient cooling
but grows by accreting mass through an accretion disc.
This second phase of star formation is called the {\it accretion phase}.
In general, the accretion disc is gravitationally unstable, and thus DF occurs 
\citep{Clark11, Greif12, Hosokawa16}.

So far, theoreticians have mostly focused on CF in the collapsing phase.
In hydrodynamics simulations, time-stepping is severely limited by the so-called
Courant conditions in very dense protostellar regions.
It is costly to follow the long-term evolution of over $10^5$ yr in
the accretion phase up to the point when the majority of protostars 
reach their zero-age main-sequence (ZAMS). 
With a semi-analytic $\alpha$-disc model, \citet{Tanaka14} 
showed that DF
is an important physical mechanism for the formation of
low-mass stars with metallicities $10^{-5}$--$10^{-3} \ \Zsun$.
More recently, with increasing computational power, several authors
successfully follow the evolution of protostellar systems in low-metallicity
accretion discs.
They report that low-mass secondary 
protostars with masses $\sim 0.1$--$1 \ \Msun$ 
form both in primordial discs \citep{Clark11, Greif12, Susa14, Hirano17, Inoue20}
and in low-metallicity discs with $10^{-6}$--$10^{-3} \ \Zsun$ \citep{Machida15, Chiaki16,
Chon21}.

Interestingly,
binary systems with iron abundances (a proxy of metallicities)
$\abH{Fe}<-3$ are discovered with recent stellar radial velocity measurements 
\citep{Arentsen19}.
\citet{Schlaufman18} report an ultra metal-poor star binary,
2MASS J18082002$-$5104378
with a metallicity ${\rm [Fe/H]} = -4.07$.
From its small binary separation (0.2 au),
we can speculate that such a system can form through DF because
the typical separation of fragments for CF is $\sim 10$ au \citep{Chiaki16}
while that for DF is $\sim 1$ au \citep{Machida15, Greif12}.

There still remain technical limitations in numerical simulations
of disc evolution.
In order to follow the long-term evolution of the accretion phase,
researchers often employ
the following two technical methods.
In the first method, called the sink particle method, dense regions are replaced with Lagrangian particles.
The mass of a sink particle is estimated from the amount of gas accreted 
into a sphere with a certain accretion radius.
In the second method, a stiff EOS is assumed in dense regions to prevent
further contraction of gas \citep{Machida15, Hirano17}.
With the sink particle technique, recent numerical studies have successfully followed the entire evolution of
accretion discs \citep{Fukushima20, Sugimura20}, but
further complicated modelling is required to reproduce the following processes that control
the multiplicity of protostars
\citep[e.g.][]{Wollenberg20}.
Mergers of protostars can efficiently reduce the number of 
protostars \citep[e.g.,][]{Susa19}.
A rapidly rotating protostar
acquiring angular momentum from accreted gas can eventually break up into two protostars
\citep{Lyttleton53, Chandrasekhar62, Chandrasekhar65, Eriguchi82}.
With the stiff EOS method, although we can follow only $\sim 10$--$100$ yr of the evolution
even with currently available computational resources
\citep{Greif12, Machida15, Hirano17}, 
this method can accurately reproduce the relevant physical processes, such as 
mergers and the break-up of protostars.

In our previous work \citepalias{Chiaki16},
we performed a series of numerical simulations of low-metallicity star formation
within cosmological  minihalos by adopting a stiff EOS technique.
We found that two distinct modes of fragmentation (CF and DF) occur
for different initial conditions.
However, in C16, with the smoothed particle hydrodynamics (SPH) scheme, 
the local Jeans length was only marginally resolved by 10 smoothing lengths.
It is known that, if a contracting gas cloud is not sufficiently resolved, spurious
fragmentation may occur
\citep{Truelove97, Truelove98}.
In addition, we followed the disc evolution only up to 50 years
after the first adiabatic core formation.
Although CF occurs along dense thin filaments in some runs,
the filaments and fragments are quickly accreted onto the most massive protostar.
In order to determine whether the filaments and fragments are 
disrupted or not, it is necessary to follow the longer evolution of the protostellar systems.

In this work, we run simulations of collapsing gas clouds and accretion discs
for two cosmological MHs with a wide range of gas metallicities $0$--$10^{-3} \ \Zsun$.
As \citetalias{Chiaki16},
we adopt metal and dust abundances and the dust size distribution
obtained from nucleosynthesis and nucleation models for a Pop III SN \citep{Umeda02, Nozawa07}.
Our numerical model considers all relevant radiative cooling processes
and chemical reactions together with grain chemistry.
Compared to \citetalias{Chiaki16}, we perform higher-resolution simulations 
to follow the fragmentation of clouds and accretion discs more accurately.
Also, we follow the evolution of protostellar systems for a longer time
of over a few hundred years using a similar stiff EOS technique.
The structure of the paper is as follows.
In Section \ref{sec:method}, we describe our numerical method and chemical models.
Section \ref{sec:results} shows the results of the simulations.
We discuss the limitations of our simulations in Section \ref{sec:discussion}
and conclude this paper in Section \ref{sec:conclusion}.


\section{Numerical models}
\label{sec:method}

We perform simulations 
using a moving-mesh/$N$-body code {\sc arepo} \citep{Springel10}.
We use two small-mass halos (minihalos; MHs) extracted from a parent cosmological simulation
of \citet{Hirano14}. 
We add heavy elements in the gas and follow their condensation
until the gas density reaches $\nH \sim 10^{16} \ \percc$.
Instead of using sink particles, we assume a stiff EOS in dense regions
to prevent further collapse and density increase.

\begin{table}
\begin{minipage}{\columnwidth}
\caption{Properties of minihalos}
\label{tab:MH}
\begin{tabular}{cccccc}
\hline
Halo & $z_{\rm form}$ & $R_{\rm vir}$ & $M_{\rm vir}$ & $M_{\rm vir,dm}$ & $M_{\rm vir,ba}$ \\
     &                     & [pc]          & [$\Msun$]     & [$\Msun$]        & [$\Msun$]        \\
\hline \hline                                                                               
A    & $15.15$        & $37.5$        & $9.7\E{4}$    & $8.1\E{4}$       &  $1.6\E{4}$      \\
B    & $20.46$        & $26.5$        & $1.6\E{5}$    & $1.4\E{5}$       &  $2.0\E{4}$      \\
\hline
\end{tabular}
\medskip \\
Note --- $z_{\rm form}$ is the formation redshifts.
$R_{\rm vir}$ and $M_{\rm vir}$ are the virial radius and mass, respectively.
$M_{\rm vir, dm}$ and $M_{\rm vir, ba}$ are the virial mass of the dark matter and baryon components, respectively.
\end{minipage}
\end{table}

\begin{table*}
\begin{minipage}{14cm}
\caption{Properties of protostellar systems}
\label{tab:PS}
\begin{tabular}{ccccccccccc}
\hline
Halo &  $Z$       & $t_{*,{\rm fin}}$ & $M_{{\rm disc}}$ & $\dot{M}_{\rm disc}$  & $T_{\rm disc}$  & $N_{*,{\rm tot}}$ & $N_{*}$ & $M_{*,{\rm tot}}$ & $\dot{M}_*$  & $T_{*}$  \\
     &  $[\Zsun]$ & [yr]              & [$\Msun$]        & [$\Msunyr$]           & [K]             &                   &         & [$\Msun$]         & [$\Msunyr$]  & [K]      \\
\hline \hline                                                                      
A    & $10^{-3}$  & 400               & $ 0.182$         & $  4.55\E{-4}$        & $   138$        &  3                & 3       & 0.521             & $1.30\E{-3}$ &  278     \\
     & $10^{-4}$  & 200               & $ 0.106$         & $  5.31\E{-4}$        & $   153$        &  3                & 3       & 0.567             & $2.83\E{-3}$ &  466     \\
     & $10^{-5}$  & 200               & $ 0.213$         & $  1.06\E{-3}$        & $   243$        &  7                & 3       & 1.03              & $5.16\E{-3}$ &  696     \\
     & $10^{-6}$  & 200               & $ 0.155$         & $  7.77\E{-4}$        & $   197$        &  5                & 3       & 1.41              & $7.04\E{-3}$ &  855     \\
     &       $0$  & 200               & $ 0.327$         & $  1.64\E{-3}$        & $   324$        & 13                & 3       & 2.60              & $1.30\E{-2}$ & 1287     \\
\hline                                                                                             
B    & $10^{-3}$  & 200               & $ 0.381$         & $  1.90\E{-3}$        & $   358$        &  2                & 2       & 1.42              & $7.11\E{-3}$ &  862     \\
     & $10^{-4}$  & 100               & $ 0.463$         & $  4.63\E{-3}$        & $   647$        & 10                & 4       & 1.88              & $1.88\E{-2}$ & 1647     \\
     & $10^{-5}$  & 100               & $ 0.373$         & $  3.73\E{-3}$        & $   560$        &  9                & 3       & 1.49              & $1.49\E{-2}$ & 1408     \\
     & $10^{-6}$  & 100               & $ 0.686$         & $  6.86\E{-3}$        & $   841$        &  4                & 1       & 5.62              & $5.62\E{-2}$ & 3416     \\
     &       $0$  & 100               & $ 0.430$         & $  4.30\E{-3}$        & $   616$        & 26                & 5       & 3.77              & $3.77\E{-2}$ & 2620     \\
\hline
\end{tabular}
\medskip \\
Note --- $M_{\rm disc}$ is the disc mass at the time $t_{*,{\rm fin}}$.
$\dot{M}_{\rm disc}$ is the average mass accretion rate onto the discs during $t_{*,{\rm fin}}$.
The temperature $T_{\rm disc}$ of the accreted gas onto the discs is calculated from 
$\dot{M}_{\rm disc} = \cs^3 /G$.
$N_{*,{\rm tot}}$ is the total number of protostars including destroyed ones during $t_{*,{\rm fin}}$.
$N_{*}$ and $M_{*,{\rm tot}}$ are the number and the total mass of protostars surviving 
at the time $t_{*,{\rm fin}}$.
$\dot {M_*}$ is the average mass accretion rate during $t_{*,{\rm fin}}$.
The temperature $T_{*}$ of the accreted gas onto the protostars is calculated from $\dot{M}_* = \cs^3 / G$.
\end{minipage}
\end{table*}

\subsection{Simulation setup}

\begin{figure}
\includegraphics[width=\columnwidth]{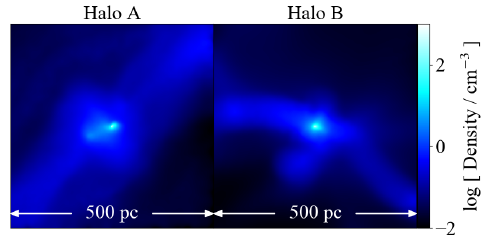}
\caption{Density-weighted density projections of Halo A and B
along the computational $z$-axis when the maximum density is $\nHmax = 10^3 \ \percc$.
Throughout this paper, we generate projected maps by integrating the contribution of
cell-generating points using a cubic spline kernel 
with a smoothing length of four times the effective cell radius (see text).}
\label{fig:MH}
\end{figure}

\begin{figure}
\includegraphics[width=\columnwidth]{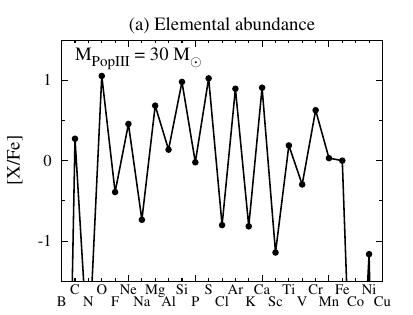}
\includegraphics[width=\columnwidth]{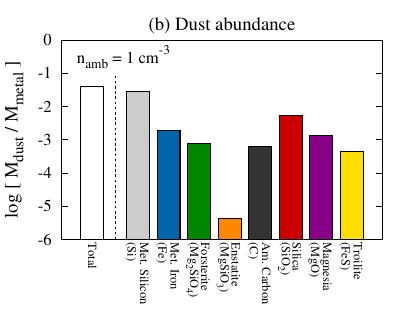}
\includegraphics[width=\columnwidth]{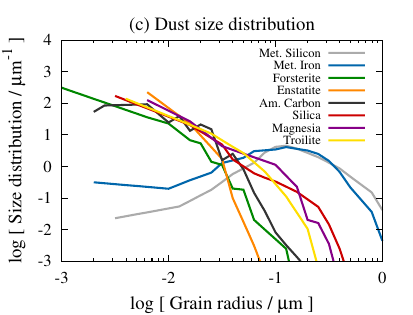}
\caption{
(a) Relative abundance $[X/{\rm Fe}]$ of each element $X$ to iron
for our Pop III SN model with mass $\MPopIII = 30 \ \Msun$.
(b) Mass fraction of dust grains (white box) and each species (colored boxes)
relative to metal for the progenitor model with an ambient gas density $\namb = 1 \ \percc$.
(c) Size distribution $\varphi _i$ of each grain species $i$ normalized to the unity.
}
\label{fig:metaldust}
\end{figure}

The parent cosmological simulation of  \citet{Hirano14}
is carried out with SPH/$N$-body simulation code {\sc gadget}-2 \citep{Springel05}.
The simulation box size is $1h^{-1}$ comoving Mpc, and
the standard $\Lambda$CDM cosmology is adopted \citep{Komatsu11}.
The regions of star-forming gas clouds are zoomed-in progressively,
and the resulting
DM and gas particle masses are $17.3h^{-1} \ \Msun$ and $3.49h^{-1} \ \Msun$, respectively.
 
A total of $\sim 100$ MHs are identified, out of which we choose two distinctive ones hereafter called
Halo A and B.\footnote{
Halo A is the same as the one used in \citet{ChiakiYoshida15}.
Halo A and B are the same as MH3 and MH1 in \citetalias{Chiaki16}, respectively.}
In \citetalias{Chiaki16}, we showed that the density structure and fragmentation
property depends not only on metallicity but also on the collapsing timescale
of clouds.
Halo A and B have respectively longest and shortest collapsing timescales $t_{\rm col} \simeq 10 t_{\rm col,0}$
and $\simeq 5 t_{\rm col,0}$ among 100 MHs investigated in \citet{Hirano14},
where $t_{\rm col,0} = 1/\sqrt{24\pi G \rho}$ is the $e$-folding time for the growth of density for an isothermal
cloud.\footnote{We show the collapse timescale in units of $t_{\rm col,0}$ instead of
the commonly used free-fall time $\tff$ because $t_{\rm col,0}$ is more appropriate for a cloud with pressure support.}
We summarize the properties of Halo A and B in Table \ref{tab:MH}.
Halo A collapses at redshift $z_{\rm form} = 15.15$, and the total mass within the 
virial radius $R_{\rm vir} = 37.5$ pc is $M_{\rm vir} = 9.7 \E{4} \ \Msun$.
Halo B forms at redshift 20.46, and
its virial radius and mass are $R_{\rm vir} = 26.5$ pc and $M_{\rm vir} = 1.6\E{5} \ \Msun$, respectively.
At the formation epoch, we cut out a cubic region around each MH with a side length of 1.15 kpc
with periodic boundaries.
Since the average and maximum gas densities in the boxes are $\sim 0.01 \ \percc$ and $\sim 100 \ \percc$, 
respectively,
the dynamical timescale in the cloud center is shorter than the average by two orders of magnitude.
Hence the dynamics at the box boundaries does not affect the cloud collapse.

To mimick a cloud enriched by a Pop III SN,
we consider the elemental abundance ratio and dust size distribution taken from a Pop III SN model of 
\citet{Nozawa07} (see Section \ref{sec:metaldust}).
We inject metals and grains uniformly in the entire simulation box for simplicity.
Note that metal abundances have deviation of $\sim 0.1$ dex
at $\sim 1$--$10$ pc from the center of enriched clouds
as recent cosmological simulations have shown \citep{Smith15, Chiaki19, Chiaki20}.

In {\sc arepo}, gravitational force is calculated with a tree method.
The gravitational softening length for
dark matter particles is fixed to 1.14 pc in all our simulations.
For the gas component, each computational cell is defined by the Voronoi diagram.
The physical quantities of a gas cell are calculated with a finite-volume method and
updated with the fluxes across the cell boundaries.
The gravity softening length of a gas cell $i$ is $2.8 \Delta x_i$,
where $\Delta x_i = (3 m_i / 4 \pi \rho _i)^{1/3}$ is the effective cell radius, and
$m_i$ and $\rho _i$ are mass and density of a cell $i$, respectively.

During the early collapsing phase, the local Jeans length progressively decreases.
It is known that spurious 
fragmentation occurs if the Jeans length 
is not sufficiently resolved \citep{Truelove97, Truelove98}.
To avoid this, we impose the so-called Jeans criterion for
cell refinement so that a local Jeans length $\ljeans$ is always resolved by more than 32 cells.
A cell which violates the criterion is divided into two cells.
Inversely, to reduce the computational cost, we derefine gas cells with
$\Delta x_i < \ljeans / 48$.
A cell which satisfies the criterion is just removed, and Voronoi mesh is reconstructed
with remaining mesh-generating points.
At a density $\nH = 10^{16} \ \percc$, the minimum size and mass of a gas cell become
$\sim 0.01$ au and $\sim 10^{-7} \ \Msun$ (0.03 Earth masses), respectively.

Note that the self-gravity of gas is smoothed out at the length beyond
the radius of a gas cell ($2.8\Delta x_i$), and this might affect the calculation of gravitational potential.
In particular, in the dense cores where we impose a stiff EOS 
(Section \ref{sec:stiffEOS}), the number
of gas cells are further reduced due to derefinement, which
might cause errors in the resulting morphology or orbital motion of protostars.

\subsection{Chemistry and cooling}

We solve all relevant chemical reactions
and radiative cooling/chemical heating processes in a self-consistent manner.
We here use a chemistry/cooling library {\sc grackle} for the usage in 
general hydro-codes \citep{Smith17}. \footnote{\url{https://grackle.readthedocs.io/}.} 
In \citet{Chiaki19}, we have developed this library by supplying with metal/dust chemistry together with
the growth of dust grains (grain growth) through accretion
of gas-phase metal molecules onto grains.
It includes large chemical reaction networks of 100 reactions of 38 gas-phase species:
$\rm e$, $\rm  H$, $\rm  H^+$, $\rm  H_2$, $\rm  H^-$, $\rm H_2^+$,
$\rm D$, $\rm  D^+$, $\rm  D^-$, $\rm  HD$, $\rm  HD^+$,
$\rm  He$, $\rm  He^+$, $\rm  He^{2+}$, $\rm HeH^+$, 
C$^+$, C, CH, CH$_2$, CO$^+$, CO, CO$_2$, O$^+$, O, OH$^+$, OH,
H$_2$O$^+$, H$_2$O, H$_3$O$^+$, O$_2^+$, O$_2$, Mg, Al, Si, SiO, SiO$_2$, S, and Fe,
and 10 grain species:
metallic silicon (Si), metallic iron (Fe), forsterite (Mg$_2$SiO$_4$),
enstatite (MgSiO$_3$), magnetite (Fe$_3$O$_4$), amorphous carbon (C),
silica (SiO$_2$), magnesia (MgO), troilite (FeS), and alumina (Al$_2$O$_3$)
with their compositions in the parentheses.
For the cells with densities $\nH > n_{\rm H,th}$, we do not update
the abundances of the chemical species.

Our chemistry model includes the collisional ionization/recombination of H, D, He, C and O.
It also includes the formation of H$_2$ through the H$^-$-process \citep{Peebles68, Hirasawa69}, 
H$_2^+$ process \citep{Saslaw67} and three-body reactions \citep{Palla83}.
We consider chemical heating and cooling associated 
with H$_2$ formation and destruction through the release and absorption of  
the binding energy of 4.48 eV per molecule, following the formulation of \citet{Hollenbach79}.
Our radiative cooling model considers ro-vibrational transition line cooling of H$_2$ and HD
as well as rotational cooling of CO, OH and H$_2$O \citep{Neufeld93, Neufeld95, Omukai10}.
We also calculate the rates of continuum cooling due to dust thermal emission \citep{Chiaki15}
and collisionally induced emission (CIE) from H$_2$ molecules \citep{Yoshida06}.

As the fraction of H$_2$ molecules increases, the specific heat ratio
$\gammaad$ for the adiabatic gas
varies from $5/3$ to the value $\gammaad ({\rm H_2})$ for the molecular gas.
At temperatures higher than $1000$ K, degrees of freedom of the
thermal motion of gas particles
are partially converted to vibrational degrees of freedom of molecules.
We use a fitting function of $\gammaad ({\rm H_2})$ as
\begin{equation}
\frac{1}{\gammaad ({\rm H_2}) - 1} = \frac{1}{2}
\left[ 5 + 2 x^2 \frac{e^x}{(e^x - 1)^2} \right],
\end{equation}
where $x = (6100 \ {\rm K})/ T$ \citep{Landau80}.
Then, we calculate $\gammaad$ as
\begin{eqnarray}
\frac{1}{\gammaad - 1} = \frac{\sum \frac{1}{\gammaad (x) - 1}n(x)}{\sum n(x)},
\end{eqnarray}
where $n(x)$ is the number density of a species $x$, and
$\gammaad (x) = 5/3$ for atoms, ions and electrons.
We only consider the contribution of primordial species because the number fraction of
other species are negligible (at most 
$1.61\E{-6} \left( Z/ 10^{-3}~\Zsun \right)$ for O atoms).
The mass fraction of each chemical species is treated as a passive scalar.
The advection of all chemical species is calculated from fluxes across each
cell interface.

We consider all relevant dust physics:
gas cooling from dust thermal emission, H$_2$ formation on grain surfaces, grain growth
and photon absorption.
The rates of these processes depend on the size distribution of grains.
We can obtain these rates from the mass fraction of grains integrated over size
under the following two approximations valid in molecular clouds. 
First, the growth rate of grains does not depend on the grain radius
$r$ when grain charge can be ignored \citep{Nozawa12}. 
Second, the number density of grain seeds is constant in molecular clouds, 
where the formation/destruction of grains is negligible \citep{Hirashita09}. 
Under the approximation, we obtain 
\begin{eqnarray}
\delta r_{i,t} ^3 + 3 \langle r \rangle _{i,0} \delta r_{i,t}^2 + 3 \langle r^2 \rangle _{i,0} \delta r_{i,t} 
+ \left[ 1 - \frac{X_t(i)}{X_0(i)} \right] \langle r^3 \rangle _{i,0} = 0.
\label{eq:grain_growth}
\end{eqnarray}
where $\langle r^n \rangle _{i,0} = \int r^n \varphi _{i,0} (r) {\rm d}r$ is the $n$'th momentum 
of the initial size distribution $\varphi _{i,0} (r)$ normalized to the unity \citep{Kozasa87}.
With this equation, we can derive the size increment $\delta r_{i,t}$ from
the mass fraction of the grain.
The initial values $\langle r^n \rangle _{i,0}$ ($n=1$--3) and $f_{ij, 0}$
are given by our Pop III SN model.
This method gives the exact solution of $\delta r_{i,t}$ 
under the above approximations in each gas cell, but
advection may cause an error due to the mixing of grains with different size distributions 
between adjacent cells.
We have confirmed that the result of three-dimensional simulations with our chemistry solver
is consistent with one-zone calculations \citep[Appendix B of][]{Chiaki19}.

The dust temperature is calculated from the balance between the rate of collisional heat
transfer from hot gas to cold grains
\begin{equation}
{\cal G} _i (r) = \pi r^2 \nH v_{\rm th} \left[ 2kT - 2k T_i (r) \right]
\label{eq:G_dust}
\end{equation}
and the rate of dust thermal emission
\begin{equation}
{\cal L} _i (r) = 4\sigma T_i(r)^4 Q_i (r) \pi r^2 \beta _{\rm cont}
\label{eq:L_dust}
\end{equation}
of the $i$-th grain species with radius $r$, where 
$v_{\rm th}$ is the thermal velocity of
gas, $\sigma$ is the Stephan-Boltzmann constant, $Q_i (r)$ is the Planck-mean of
the grain absorption coefficient, and $\beta _{\rm cont}$ is the escape fraction of
continuum radiation.
We calculate the rate of gas cooling as $n_i \int {\cal G} _i (r) \varphi _{i} (r, t) {\rm d}r$.





\subsection{Stiff EOS}
\label{sec:stiffEOS}

The computational timestep decreases with increasing density.
To follow the evolution of an accretion disc
after optically-thick cores (protostars) form at densities $\nH \gtrsim 10^{16} \ \percc$, 
we prevent density increase by adopting a stiff EOS technique.
We force the gas to follow a polytropic EOS $p \propto \rho ^{\gammaeff}$
at densities above $n_{\rm H,th} = 10^{16} \ \percc$
instead of solving chemistry/cooling.
We set
\begin{equation}
 \gammaeff =\begin{cases}
    1.4 + 5.2 (\log \nH -16) &  \text{for}~16 < \log \nH < 16.5 \\
    4.0 & \text{for}~\log \nH > 16.5
  \end{cases}
 \label{eq:EOS}
\end{equation}
so that $\gammaeff$ increases gradually from the adiabatic index 1.4 of the
molecular gas to 4.0. 
The choice of the value 4.0 is motivated from a particular EOS that
reproduces the mass-radius relation of protostars \citep{Machida15}.
Hereafter we call a region with densities $\nH > n_{\rm H,th}$ as a ``protostar''.\footnote{
In reality, at higher densities $\nH \sim 10^{16}$--$10^{18} \ \percc$, gas thermal energy
is partly consumed to H$_2$ dissociation.
This chemical gas cooling induces second collapse until all H$_2$ molecules are consumed.
At that time the gas again collapses adiabatically and a hydrostatic core forms.
The core hosted by the first core is defined as a protostar \citep{Larson69}.}

Here, we use the time $t_*$ to measure the evolution of a protostellar system.
The origin of $t_*$ is the time when the maximum density $\nHmax$ among gas cells first 
reaches $ > n_{\rm H,th}$.
We terminate our simulations at $t_{*, {\rm fin}} = 100$--$400$ yr depending on the runs (see Table \ref{tab:PS}).
We follow longer evolution for larger metallicity and for Halo A because its accretion
timescale is longer than Halo B.

\subsection{Cloud fragmentation vs. Disk fragmentation}
\label{sec:CF_DF}
We can expect that the two fragmentation modes, CF and DF, occur
in our simulations as we found in \citetalias{Chiaki16}.
To discriminate one from the other, 
we estimate the orbital rotational velocity of each protostar with respect to the center of the mass of the system, $\fkep = v_{\rm rot} / v_{\rm Kep}$, in units of Kepler velocity
$v_{\rm Kep} = \sqrt{G M_{*, {\rm tot}}/R}$ at the formation time of a protostar, where
$M_{*, {\rm tot}}$ is the total mass of protostars, and $R$ is the distance between the protostar
to the center of mass of the protostellar system.
Generally, CF occurs in a cloud envelope with $\fkep \sim 0.5$ \citep{Hirano14} 
while DF occurs in a rotationally supported disc with $\fkep \sim 1$.
Therefore, we regard that the protostar forms through CF if
$\fkep < 0.75$ or DF otherwise.

 \begin{figure*}
 \includegraphics[width=17.0cm]{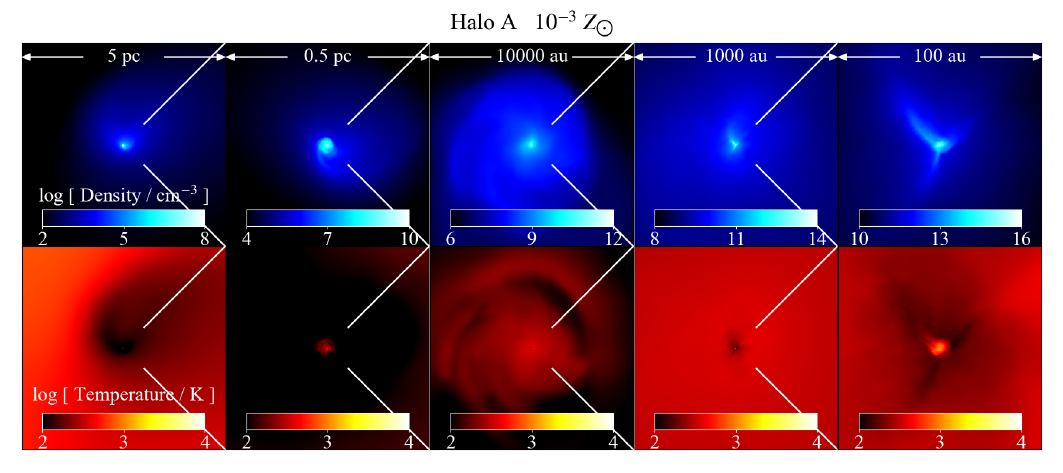}
 \includegraphics[width=17.0cm]{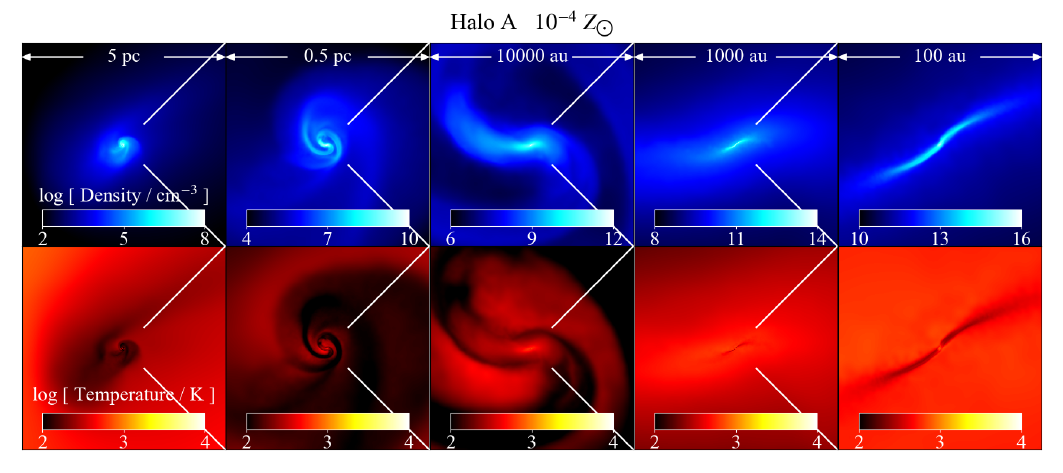}
 \includegraphics[width=17.0cm]{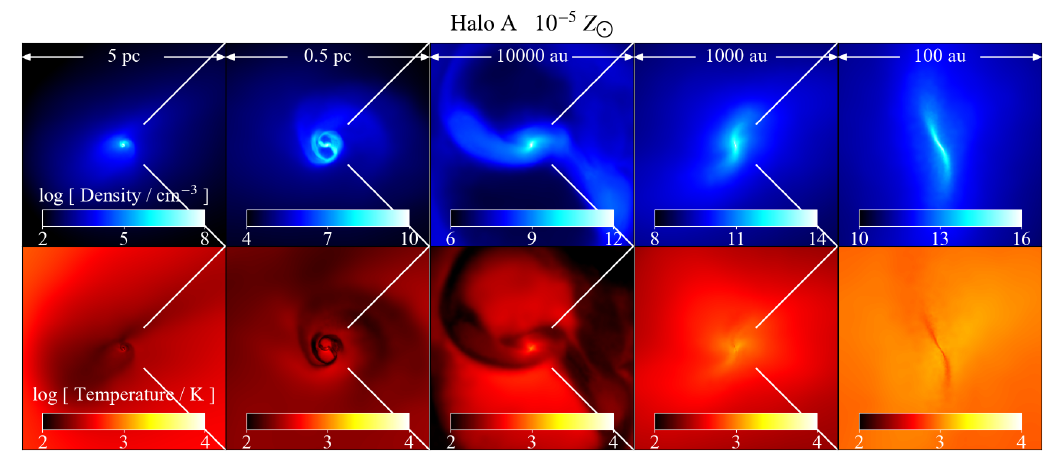}
 \caption{
Projections of density and temperature for Halo A
when the maximum gas density reaches $10^{16} \ \percc$
along the computational $z$-axis.
We plot the sequential zoom-in from 5 pc (leftmost panel) to 
100 au (rightmost panel)
for metallicities 0--$10^{-3} \ \Zsun$ from bottom to top.
 }
 \label{fig:snapshots_col_H0522_nH}
 \end{figure*}
\addtocounter{figure}{-1}
 \begin{figure*}
 \includegraphics[width=17.0cm]{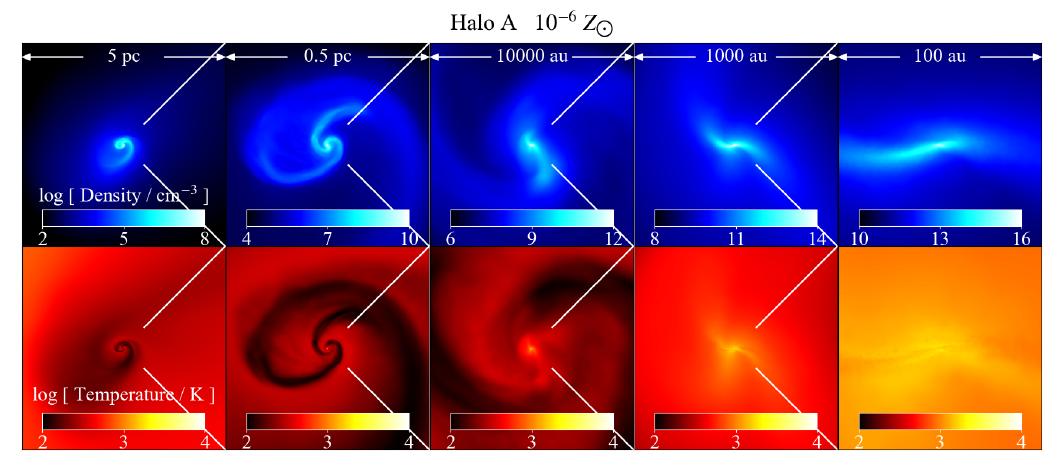}
 \includegraphics[width=17.0cm]{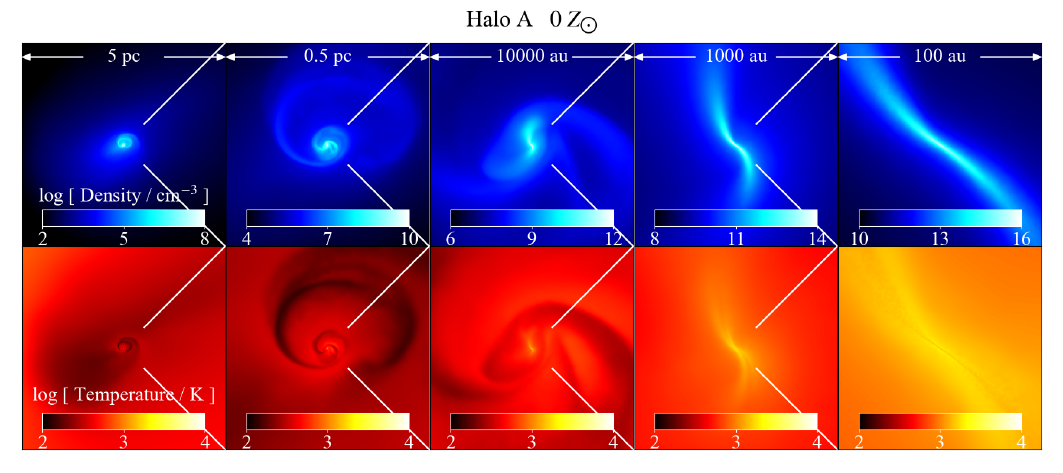}
 \caption{
 {\it cont.}
 }
 
 \end{figure*}

 \begin{figure*}
 \includegraphics[width=17.0cm]{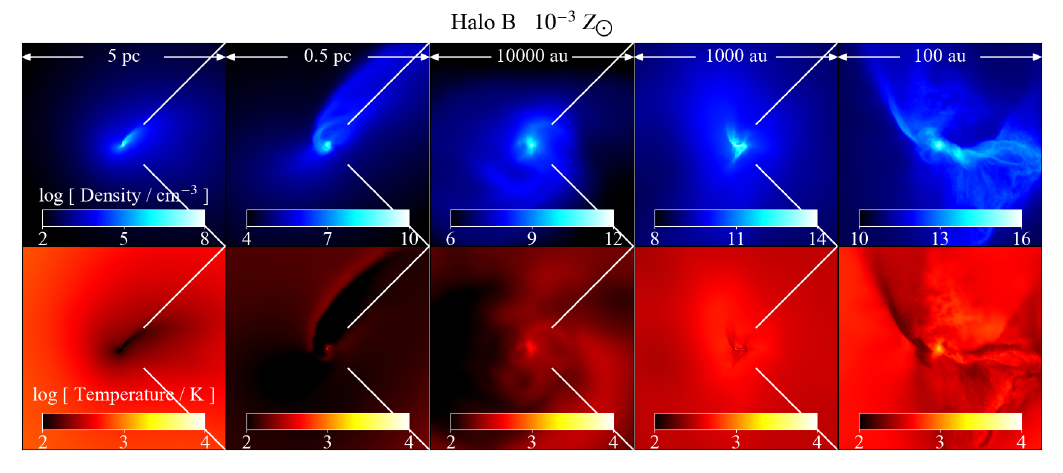}
 \includegraphics[width=17.0cm]{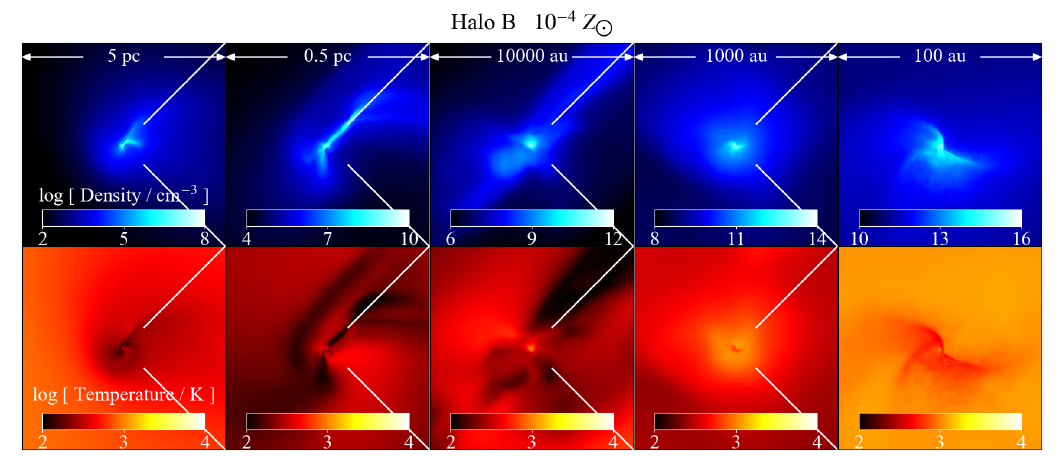}
 \includegraphics[width=17.0cm]{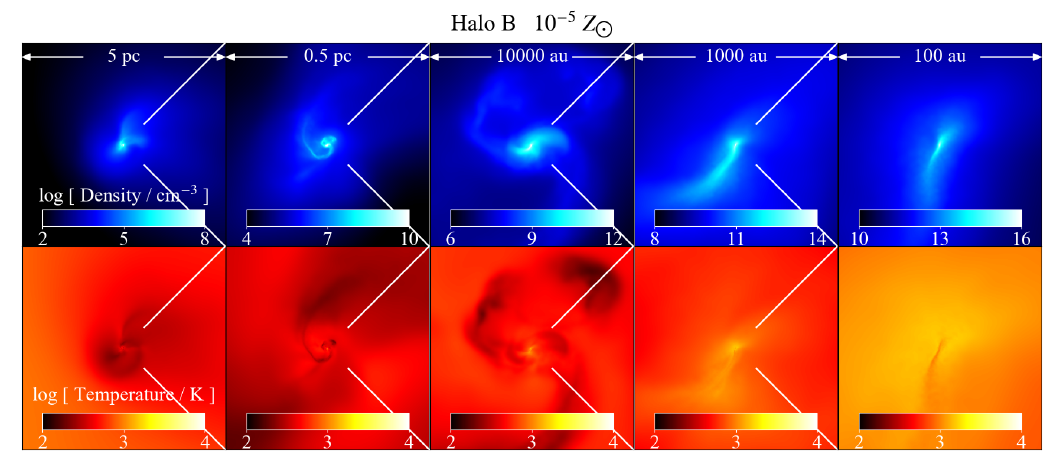}
 \caption{
 Same as Fig. \ref{fig:snapshots_col_H0522_nH} but for Halo B.
 }
 \label{fig:snapshots_col_H0611_nH}
 \end{figure*}
\addtocounter{figure}{-1}
 \begin{figure*}
 \includegraphics[width=17.0cm]{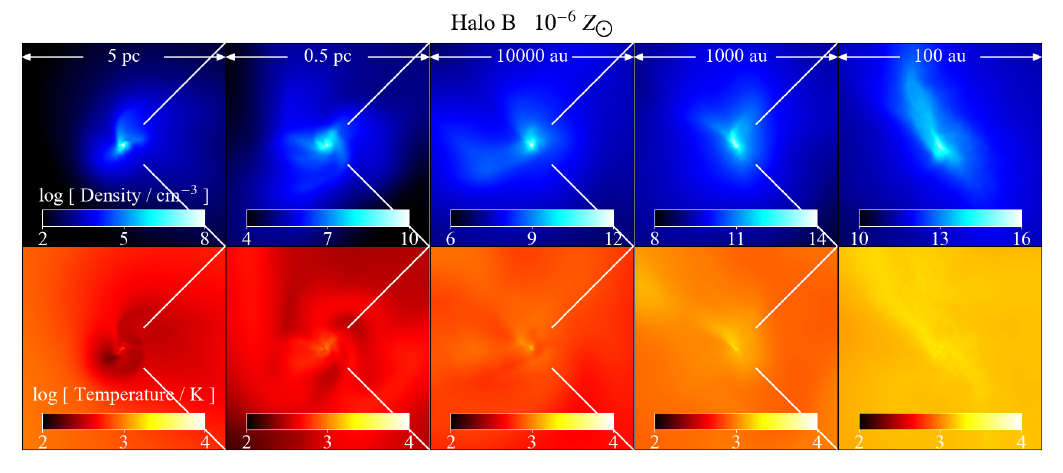}
 \includegraphics[width=17.0cm]{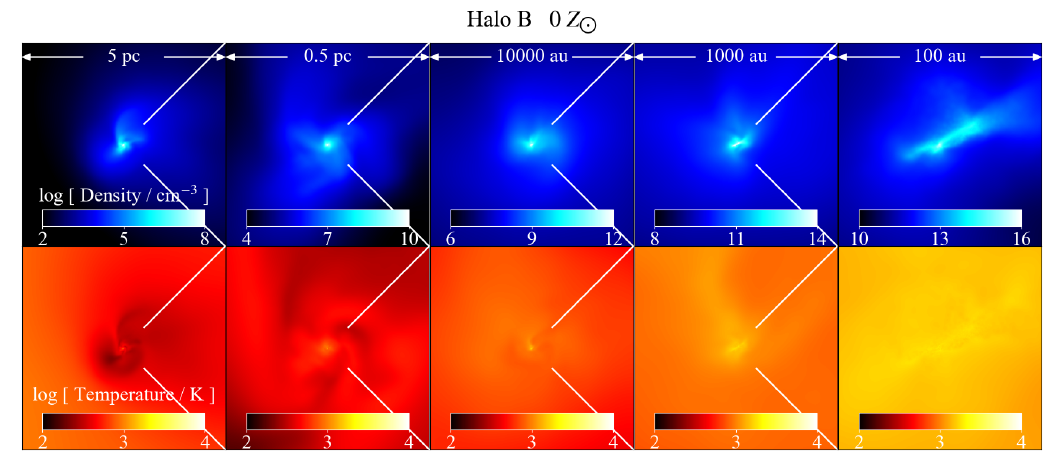}
 \caption{
  {\it cont.}
 }

 \end{figure*}

\begin{figure*}
 \includegraphics[width=18.0cm]{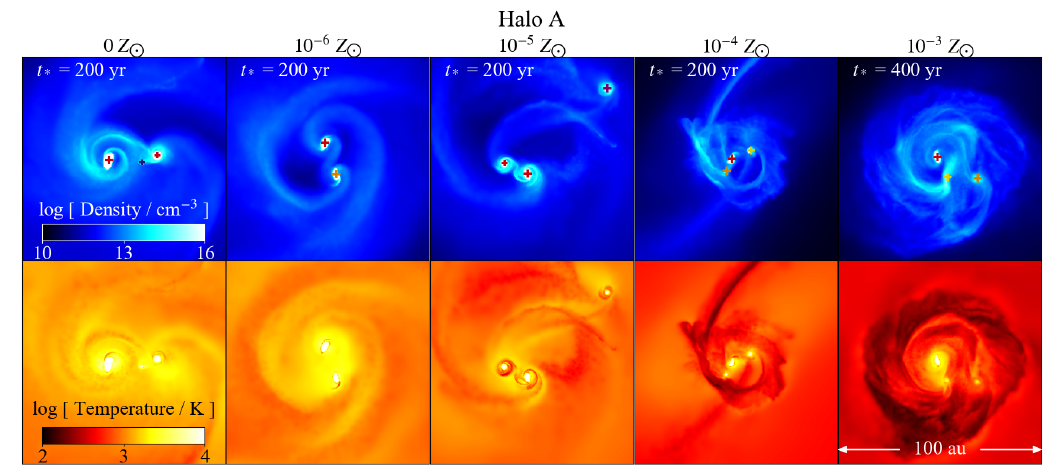}
\caption{Projections of density (top) and temperature (bottom)
in accretion discs when we terminate the simulations
for Halo A with metallicities $0$--$10^{-3} \ \Zsun$ from left to right.
The cross symbols indicate the center of mass of protostars.
We plot the results along the total angular momentum vector of the region 
with densities $> 10^{14} \ \percc$.
}
\label{fig:snapshots_acc_H0522_nH}
\end{figure*}

\begin{figure*}
 \includegraphics[width=18.0cm]{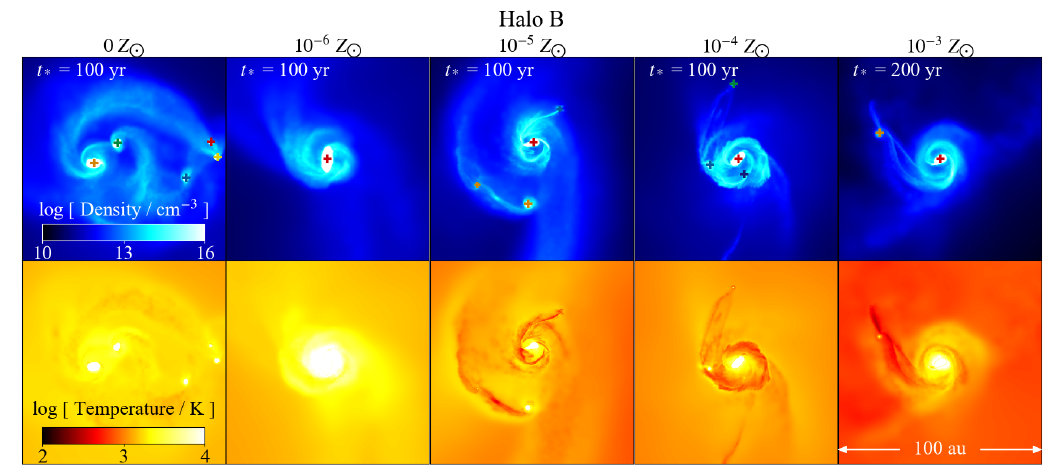}
\caption{Same as Fig. \ref{fig:snapshots_acc_H0522_nH} but for Halo B.}
\label{fig:snapshots_acc_H0611_nH}
\end{figure*}

\subsection{Initial metal/dust properties}
\label{sec:metaldust}

We adopt realistic metal/dust properties (abundances of heavy elements and
grain species and dust size distribution) taken from nucleosynthesis and nucleation models of a Pop III SN
\citep{Umeda02, Nozawa07}.
We assume a progenitor mass $\MPopIII = 30 \ \Msun$
motivated by recent simulations.
The local maximum of Pop III stellar mass distribution is at $30 \ \Msun$ \citep{Susa14, Hirano14}.
Also, the elemental abundance ratios of metal-poor stars best fit with  
progenitor models
with $\MPopIII = 25$--$30 \ \Msun$ \citep{Ishigaki18}.

Fig. \ref{fig:metaldust}a shows the abundances of heavy elements in total of 
the gas and solid phases.
This model reproduces the enhancement of $\alpha$ elements such as O, Mg, and Si,
of EMP stars although ${\rm [\alpha /Fe]}$ is higher than the typical value of EMP stars ($\sim 0.4$).
In this model, the oxygen abundance relative to iron is ${\rm [O/Fe]} = 1.05$.
We expect that radiative cooling by O-bearing molecules OH and H$_2$O affect the 
cloud fragmentation as suggested by \citetalias{Chiaki16}.
From the relative abundance of carbon to iron ${\rm [C/Fe]} = 0.274$,
the stars will be classified to C-normal stars by a definition of C-enhanced
metal-poor (CEMP) stars \citep[${\rm [C/Fe]} > 0.7$;][]{Aoki07}.

Fig. \ref{fig:metaldust}b and c show the initial abundances and size distribution
of grain species, respectively, in our progenitor model.
Typically, grains form in the expanding ejecta a few hundred days after the SN explosion \citep{Nozawa03}.
All refractory elements such as Mg, Si, and Fe are locked up into grains
as in the local ISM.
At $\sim 10^4$ yr after the explosion, reverse shocks start propagating backwards into ejecta.
Between the forward and reverse shocks,
high energy ions sputter the newly forming grains, and the mass fraction of
metals locked up into grains decreases.
The efficiency of grain destruction depends on the ambient gas density $\namb$
\citep{Nozawa07, Bianchi07}.
We here consider the grain destruction model with $\namb = 1 \ \percc$ because the
typical gas density of an H {\sc ii} region created by a progenitor Pop III star
is $\sim 0.1$--$1 \ \percc$ \citep{Kitayama04, Whalen04, Whalen08}.
The dust-to-metal mass ratio
is smaller in this model (3.95\%) than in the local ISM $\simeq 50$\% \citep{Pollack94}.

The absolute amount of metals is given by 
the total mass fraction of all elements heavier than He.
For a metallicity $Z$, the carbon and iron abundances in our progenitor model are
\begin{eqnarray}
A({\rm C}) &=&  
4.05 + \log \left( \frac{Z}{10^{-4} \ \Zsun} \right), \\
{\rm [Fe/H]} 
&=& A({\rm Fe}) - A_{\bigodot} ({\rm Fe}) \nonumber \\
&=& -4.66 + \log \left( \frac{Z}{10^{-4} \ \Zsun} \right).
\label{eq:FeH}
\end{eqnarray}
The abundance $A(X)$ of an element $X$ is calculated as
\begin{equation}
A(X) = 12 + \log \left( \frac{ M_X }{ M_{\rm met} } \frac{Z }{\mu _X \XH} \right),
\end{equation}
where $M_X$ and $M_{\rm met}$ are mass of $X$ and metals ejected from the SN, respectively,
and $\mu _X$ is molecular weight of element $X$ ($\mu _{\rm C} = 12$ and $\mu _{\rm Fe} = 56$).
We assume the primordial hydrogen mass fraction $\XH = 0.76$
and give metallicities in units of solar metallicity $Z_{\bigodot} = 0.02$.

Hereafter, our runs with different halos and metallicities
are described as the combination of ids of halos \{{\tt HA}, {\tt HB}\}
and metallicities \{{\tt Z0}, {\tt Z-6}, {\tt Z-5}, {\tt Z-4}, {\tt Z-3}\}.
For example, the run {\tt HAZ-4} indicates the run for Halo A and a metallicity
$10^{-4} \ \Zsun$.


\section{Results}
\label{sec:results}

\subsection{Overview}

It has been conventionally accepted that dust cooling induces CF, and low-mass stars
can form through competitive accretion among the fragments at metallicities
$\gtrsim 10^{-5} \ \Zsun$ \citep[e.g.,][]{Omukai00}.
Figs. \ref{fig:snapshots_col_H0522_nH} and \ref{fig:snapshots_col_H0611_nH} show
the density-weighted projections of density and temperature when $\nHmax$ first
reaches $n_{\rm H,th}$ for Halo A and B,
respectively.
dust cooling induces CF for $10^{-3} \ \Zsun$, but the fragments are accreted by the central
most massive protostar within 50 yr.
For $10^{-5}$ and $10^{-4} \ \Zsun$, even with efficient dust cooling,
CF does not occur.
At extremely low metallicities of $0$ and $10^{-6} \ \Zsun$, dense filaments form and fragment within $\sim 20$ yr 
owing to H$_2$
cooling, in a different manner from the conventional notion.
Although H$_2$ cooling is less efficient than dust cooling, 
the filaments can become sufficiently massive to be unstable because the accretion rate is larger 
($> 0.03 \ \Msunyr$) in the hotter gas ($\sim 1000$ K) than in a cool, metal-poor gas.
However, the fragments on the filaments are accreted onto the primary protostar
within $\sim 50$ yr.
This suggests that CF is a ``transitional'' event, and does not determine the final state of the
cloud core. 

Figs. \ref{fig:snapshots_acc_H0522_nH} and \ref{fig:snapshots_acc_H0611_nH} show
the density-weighted projections of density and temperature when we terminate
the simulations at the time $t_{*, {\rm fin}}$ after the formatiothe first protostar for Halo A and B,
respectively.
In all runs but for {\tt HBZ-6}, 
accretion discs form at $t_* \sim 10$ yr, when the sufficient amount of 
ambient gas is accreted onto the cores.
In the sufficiently massive discs, DF eventually occurs as seen in
Figs. \ref{fig:snapshots_acc_H0522_nH} and \ref{fig:snapshots_acc_H0611_nH}.
We find that DF is the main process of fragmentation {\it regardless of metallicity}.

In the discs, as many as 26 protostars form, some of which are destroyed within a few 100 yr.
The primary protostar grows through gas accretion while secondary protostars form through 
the break-up of their parent protostar after it strongly deforms by the centrifugal force, or through the
interaction of spiral arms around a massive protostar.
Some protostars are destroyed through merger with another protostar.
When we terminate our simulations at 100--400 yr after primary protostar formation,
1--5 protostars survive. 
Table \ref{tab:PS} shows the total mass of protostellar system $M_{*,{\rm tot}}$. 
The mass is $0.52$--$3.8 \ \Msun$ depending on the mass accretion rate 
and metallicity (Section \ref{sec:accretion}).

\begin{figure*}
 \includegraphics[width=\textwidth]{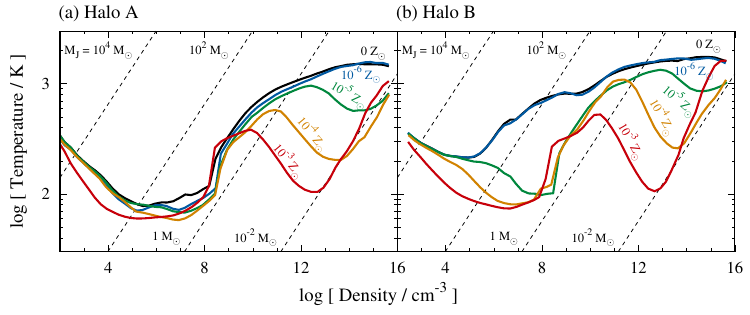}
\caption{
{\it Coloured solid curves}:
temperature evolution of cloud cores for (a) Halo A and (b) Halo B with metallicities 
$0$ (black), $10^{-6} \ \Zsun$ (blue), $10^{-5} \ \Zsun$ (green), $10^{-4} \ \Zsun$ (orange) and $10^{-3} \ \Zsun$ (red).
In each run, we dump snapshots at every time when $\nHmax$ increases by $0.25$ dex, where
$\nHmax$ is the maximum density in each snapshot.
At each output time, we plot the volume-weighted average of density
and the mass-weighted average of temperature in the core of the clouds
on the horizontal and vertical axes, respectively.
In this analysis, the cloud core is defined as a region with densities above $\nHmax / 3$.
{\it Black dotted lines}: Jeans mass at the corresponding density and temperature.}
\label{fig:nT}
\end{figure*}

\begin{figure*}
 \includegraphics[width=14cm]{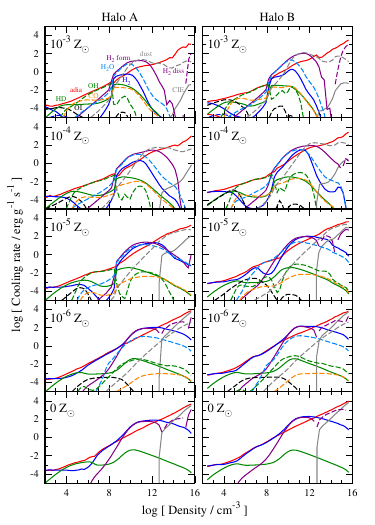}
\caption{
Evolution of cooling/heating rates as a function of mean
density of the cloud cores
for Halo A (left panels) and B (right panels) with metallicities
0--$10^{-3} \ \Zsun$ from bottom to top.
We plot the rates of adiabatic heating (red solid) and 
radiative cooling of H$_2$ (blue solid), HD (green solid),
O {\sc i} (black dashed), CO (orange dashed), OH (green dashed), H$_2$O (cyan dashed),
dust (grey dashed) and CIE (grey solid curve).
We also plot the chemical cooling/heating:
H$_2$ formation cooling (purple solid) and dissociation heating (purple dashed).
}
\label{fig:nL}
\end{figure*}

\begin{figure*}
\includegraphics[width=\textwidth]{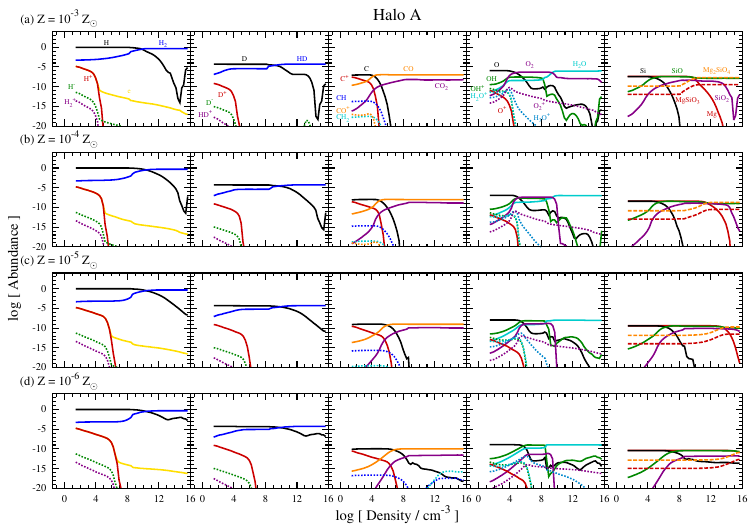}
\caption{
Evolution of species abundances relative to hydrogen nuclei as a function of mean
density of the cloud cores
for Halo A with metallicites 
(a) $10^{-3} \ \Zsun$, (b) $10^{-4} \ \Zsun$, (c) $10^{-5} \ \Zsun$, and (d) $10^{-6} \ \Zsun$.
For silicate grains (Mg$_2$SiO$_4$ and MgSiO$_3$), we show the number fraction of Si nuclei locked up
into the grains.
}
\label{fig:ny_H0522}
\end{figure*}

\begin{figure*}
\includegraphics[width=\textwidth]{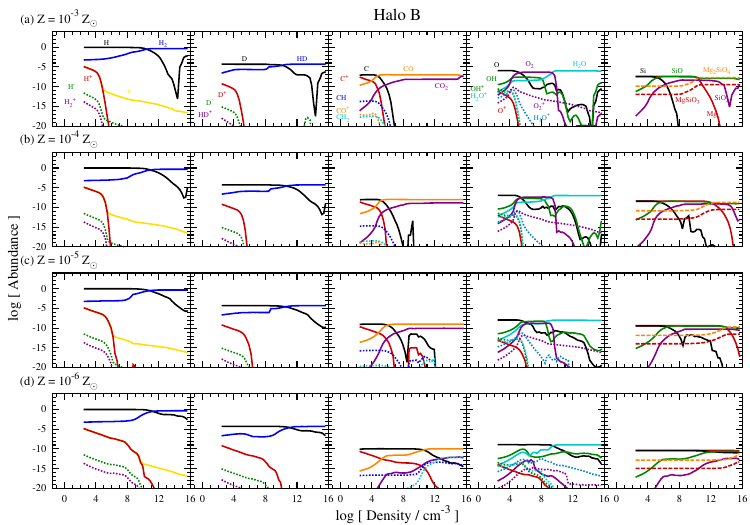}
\caption{
Same as Fig. \ref{fig:ny_H0522} but for minihalo Halo B.
}
\label{fig:ny_H0611}
\end{figure*}

\subsection{Collapsing phase}
\label{sec:collapse}

In this section, we examine the thermal evolution of our clouds in more detail, 
and then define the condition under which CF occurs.

\subsubsection{Thermal evolution}

A gas cloud collapses and condenses toward the center of the gravitational potential generated by the host dark matter halo.
Since cloud deformation and fragmentation are controlled by the thermal evolution of the gas,
we first discuss cooling/heating processes and chemical evolution in our clouds.
Figs. \ref{fig:nT} and \ref{fig:nL} show the evolution of the
temperature and cooling rates as a function of the average density in the cloud core.
Figs. \ref{fig:ny_H0522} and \ref{fig:ny_H0611} show the evolution of spieces abundances
for Halo A and B, respectively.
The temperature evolution is different between the two MHs even with the same metallicity.
Since the collapsing timescale $\tcol = \rho / \dot \rho$ is longer for Halo A than for Halo B
by a factor of two, the gas temperature is lower with the smaller compressional heating rate.
The evolution of chemical species abundances does not significantly differ
between the two halos with a fixed metallicity (see Figs \ref{fig:ny_H0522} and \ref{fig:ny_H0611})
but the thermal evolution notably or significantly differs between the runs with different gas metallicities.

The overall thermal evolution can be described as follows.
At low densities $\nH \lesssim 1 \ \percc$, the gas temperature increases adiabatically,
because any cooling process does not work efficiently.
The hydrogen molecule fraction increases up to $y({\rm H_2}) \sim 10^{-3}$ through the H$^-$-process as
\begin{eqnarray}
{\rm H} + {\rm e^-} &\to& {\rm H^-} + \gamma , \nonumber \\
{\rm H^-} + {\rm H} &\to& {\rm H_2} + {\rm e^-}, \nonumber
\end{eqnarray}
catalyzed by free electrons. Then, the gas temperature declines through hydrogen molecular cooling.
The molecular fraction increases as metallicity increases because of more efficient molecular formation 
on grain surfaces.
In models where the temperature drops below 150 K, HD molecules form through the reaction
\begin{eqnarray}
{\rm D} + {\rm H_2} &\to& {\rm HD} + {\rm H}, \nonumber 
\end{eqnarray}
and rotational transition line cooling of HD molecules becomes important.

Metal atoms/ions or molecules also contribute to gas cooling for metallicities
$\gtrsim 10^{-5} \ \Zsun$.
With a metallicity $10^{-3} \ \Zsun$, fine-structure cooling of O {\sc i} and rotational 
transition line cooling of CO become dominant at $\nH \sim 10^2$--$10^4 \ \percc$ 
and $10^4$--$10^6 \ \percc$, respectively.
Since O is dominant over C ($y_{\rm O} / y_{\rm C} > 11$) in our progenitor model, 
C {\sc i} and C {\sc ii} fine-structure cooling is inefficient.
For metallicities above $10^{-5} \ \Zsun$, OH molecules form through
\begin{eqnarray}
{\rm O} + {\rm H} &\to& {\rm OH} + \gamma , \nonumber
\end{eqnarray}
and OH cooling becomes dominant.
With these molecular cooling, gas temperature reaches the temperature of
the cosmic microwave background (CMB), 44 K and 59 K 
for Halo A and B, respectively, and temperature floor appears in Fig. \ref{fig:nT}.
The nearly isothermal evolution induces cloud deformation \citep{Tsuribe06, Sugimura17}.
The second and third columns of Fig. \ref{fig:snapshots_col_H0522_nH} shows dense ($\nH \sim 10^5$--$10^8 \ \percc$)
and cool ($T \simeq 60$ K) spiral arms.

The cloud deformation ceases at density $\nH \sim 10^8 \ \percc$, where efficient
gas heating occurs due to the release of binding energy 
along with hydrogen molecule formation through three-body reactions:
\begin{eqnarray}
{\rm H} + {\rm H} + {\rm H}   &\to& {\rm H_2} + {\rm H} , \nonumber \\
{\rm H} + {\rm H} + {\rm H_2} &\to& {\rm H_2} + {\rm H_2}. \nonumber
\end{eqnarray}
The gas cloud again collapses stably against deformation, and quasi-spherical
hydrostatic cores form (the forth column of Figs. \ref{fig:snapshots_col_H0522_nH} and
\ref{fig:snapshots_col_H0611_nH}).
The ambient gas is accreted onto the core,
and thus the mass and density of the core can continue to
increase.

\begin{figure*}
\includegraphics[width=15cm]{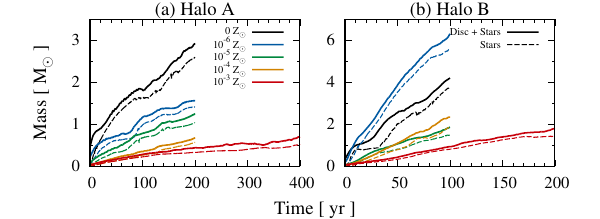}
\caption{
Total mass of accretion discs and protostars (regions with densities $>10^{14} \ \percc$; solid) 
and only protostars (regions with densities $> 10^{16} \ \percc$; dashed) as a function of time
for metallicities
$0$ (black), $10^{-6} \ \Zsun$ (blue), $10^{-5} \ \Zsun$ (green), $10^{-4} \ \Zsun$ (orange)
and $10^{-3} \ \Zsun$ (red)
for Halo A (left column) and B (right column).
}
\label{fig:tmtot}
\end{figure*}

\begin{figure*}
\includegraphics[width=0.99\textwidth]{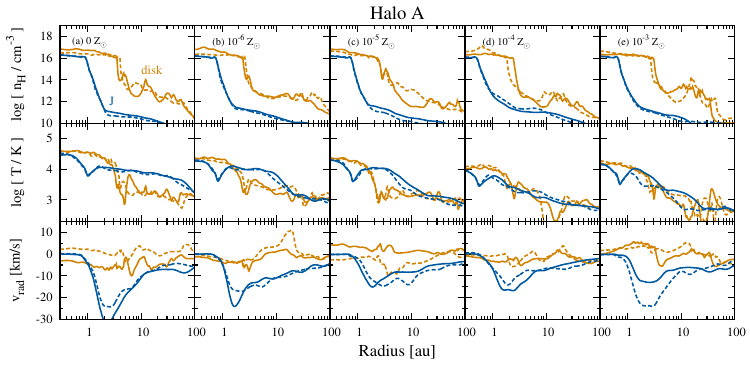}
\caption{
Density $\nH$ (top panels), temperature $T$ (middle panels) and radial velocity $v_{\rm rad}$ 
(bottom panels) as a function of distance from the center of mass of the most massive protostar.
We plot the results for metallicites 
(a) $0 \ \Zsun$, (b) $10^{-6} \ \Zsun$, (c) $10^{-5} \ \Zsun$, (d) $10^{-4} \ \Zsun$, and (e) $10^{-3} \ \Zsun$
for Halo A at the time when we terminate the simulations.
The blue and orange curves show the results along the axes
parallel and perpendicular (vertical axis in Fig. \ref{fig:snapshots_acc_H0522_nH}) to the angular momentum vector $\vb{J}$.
The solid and dashed curves depict the opposite directions along these axes.
}
\label{fig:rn_H0522}
\end{figure*}

\begin{figure*}
\includegraphics[width=0.99\textwidth]{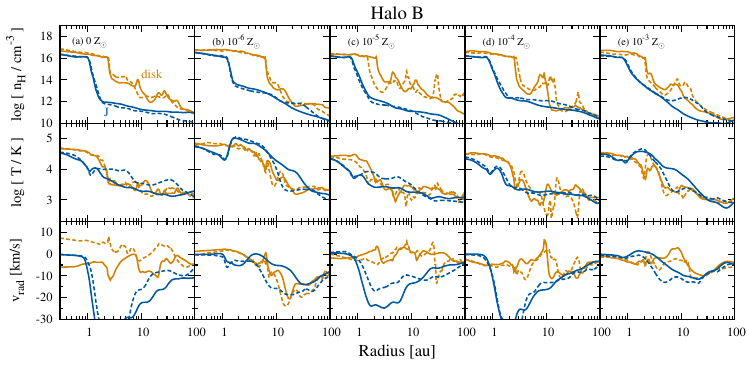}
\caption{
Same as Fig. \ref{fig:rn_H0522} but for a minihalo Halo B.
}
\label{fig:rn_H0611}
\end{figure*}

At $\nH \sim 10^{10} (Z/10^{-4} \ \Zsun) ^{-2} \ \percc$, silicate grains start 
growing by accreting Mg atoms, SiO molecules, and H$_2$O molecules in the gas phase.
Grain growth proceeds until all Mg atoms are eventually depleted 
(see Figs. \ref{fig:ny_H0522} and \ref{fig:ny_H0611}).
The grain radius most rapidly increases at densities 
$n_{\rm H, gg} \sim 10^{11} (Z/10^{-4} \ \Zsun) ^{-2} \ \percc$.
This is consistent with the results of one-zone calculations
\citep[see fig. 8 of][]{Chiaki15} and a simple estimate of $n_{\rm H, gg}$
from the balance between the growth and dynamical timescales:
\begin{eqnarray}
n_{\rm H, gg} &=& 4.0\E{11}~\percc
\left( \frac{T}{1000~{\rm K}} \right)^{-1}
\left( \frac{Z}{10^{-4}~\Zsun} \right)^{-2}.
\end{eqnarray}
As the grain radius monotonically increases,
the mass fraction of grains to its constituent elements increases (Eq. \ref{eq:grain_growth}).
The upper limit of the increment $\delta r _i (t)$ of the grain radius
does not depend on the metallicity because it is determined by the initial
mass fraction of grains with respect to its available constituent elements.
For the dominant species, forsterite (Mg$_2$SiO$_4$), the initial mass fractions of
grains and Mg to metals are $7.73\E{-4}$ and $3.87\E{-2}$, respectively.
The integrated momenta of the size distribution are
($\langle r \rangle _{\Forsterite, 0})$, $\langle r^2 \rangle _{\Forsterite, 0})$, 
$\langle r^3 \rangle _{\Forsterite, 0})$) = ($4.70\E{-3} \ \um$, $6.29\E{-5} \ \um ^2$, 
$1.71\E{-6} \ \um^3$).
From Eq. (\ref{eq:grain_growth}), $\delta r_{\Forsterite} = 5.76\E{-2} \ \um$ 
when all Mg is depleted.
For metalliciteis $>10^{-5} \ \Zsun$, $\delta r_{\Forsterite}$ reaches this value.
For $10^{-6} \ \Zsun$, Mg is partly accreted onto the grains, and
$\delta r_{\Forsterite} = 4.98\E{-2}$ and $2.83\E{-2} \ \um$
for Halo A and B, respectively, at a density $10^{16} \ \percc$.
The gas cooling rate due to dust thermal emission is enhanced by grain growth,
and the temperature decreases for metallicities above $10^{-5} \ \Zsun$.
Dust cooling enhances cloud deformation, and the filamentary structure develops
(the fifth column of Figs. \ref{fig:snapshots_col_H0522_nH} and \ref{fig:snapshots_col_H0611_nH}).
At $\nH \sim 10^{13} (Z/10^{-4} \ \Zsun) ^{-1} \ \percc$,
the gas and grains are thermally coupled, and the gas becomes optically thick
in continuum. Then, the temperature increases approximately adiabatically
(Eqs. \ref{eq:G_dust} and \ref{eq:L_dust}).
In this region, the gas hardly collapses, and a hydrostatic core (protostar) forms.

\subsubsection{Fragmentation of clouds}
\label{sec:CF}

We identify protostars forming through CF in each run,
following the criterion described in Section \ref{sec:CF_DF}.
CF does not occur for metallicities $10^{-5}$--$10^{-4} \ \Zsun$ even though
dust cooling is efficient at densities $10^{14}$--$10^{16} \ \percc$ (Fig. \ref{fig:nT}).
This can be explained as follows:
H$_2$ formation heating once stabilizes the contracting gas at densities
$10^{8}$--$10^{11} \ \percc$ (Fig. \ref{fig:nT}).
The cloud core becomes spherical when the dust cooling begins to operate, and the timescale
for deformation is longer than the collapse timescale 
as found by \citet{Tsuribe08} and \citetalias{Chiaki16}.
Although we use the same cloud as \citetalias{Chiaki16},
we obtain the different outcome.
For Halo B (MH1 in \citetalias{Chiaki16}) with a metallicity $10^{-4} \ \Zsun$,
CF does not occur in this work, but CF does occur in \citetalias{Chiaki16}.
This is due to the numerical resolution as we shall discuss in more detail in 
Section \ref{sec:resolution}.

For {\tt HBZ-3}, cloud deformation is enhanced by dust cooling,
and knotty filamentary structures develop in a manner termed ``filament fragmentation'' in \citetalias{Chiaki16}.
The whole filaments are accreted onto PS1 
by $t_* = 50$ yr, before the density
perturbations grow gravitationally.
Similarly, for {\tt HAZ0} and {\tt HBZ-6}, filament fragmentation is promoted at $t_* \sim 20$ yr 
by the combined effect of H$_2$ cooling and the high initial accretion rate 
($\dot M_* > 0.03 \ \Msunyr$), yielding
six and four protostars, respectively.
For {\tt HAZ0} and {\tt HBZ-6}, $\fkep$ of the protostars is $\leq 0.58$ and $\leq 0.43$, respectively.
This indicates that fragmentation occurs before the formation of an accretion disc in these runs.
We find that protostars can form through CF, but
all or most of them are accreted onto PS1 within a few yr
in the subsequent accretion phase.

\begin{figure*}
\includegraphics[width=\textwidth]{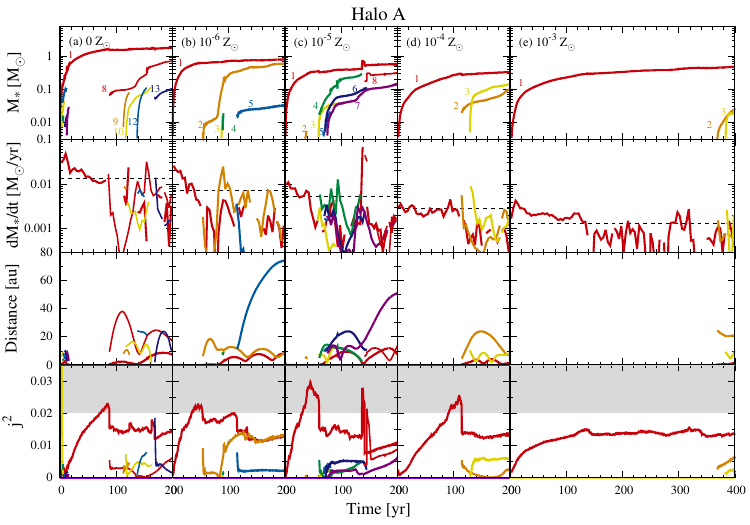}
\caption{
Time evolution of protostellar mass $M_*$, mass accretion rate $\dot M_*$, 
distance from the center of mass of the protostellar system, and 
the ratio of rotational energy to gravitational energy $j^2$ (Eq. \ref{eq:j2}).
We plot the results with metallicities
(a) $0$, (b) $10^{-6} \ \Zsun$, (c) $10^{-5} \ \Zsun$, (d) $10^{-4} \ \Zsun$, and (e) $10^{-3} \ \Zsun$
for Halo A.
The black dashed lines in the second row represent the average mass accretion rate.
The grey shaded area represents the region of $j^2$ where break-up of a protostar
occurs (see text).
}
\label{fig:tm_H0522}
\end{figure*}

\begin{figure*}
\includegraphics[width=\textwidth]{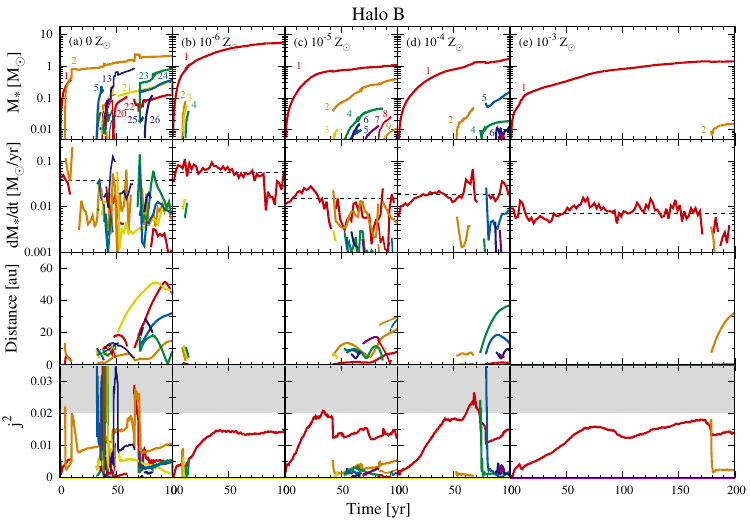}
\caption{
Same as Fig. \ref{fig:tm_H0522} but for Halo B.
The accretion rate is calculated every 2 yr.
}
\label{fig:tm_H0611}
\end{figure*}

\subsection{Accretion phase}
\label{sec:accretion}

\subsubsection{Accretion discs}

In all cases, a rotationally supported disc forms quickly after PS1 accretes 
a sufficient amount of gas with finite angular momentum
(Figs. \ref{fig:snapshots_acc_H0522_nH} and \ref{fig:snapshots_acc_H0611_nH}).
In this study, we define a circumstellar disc as the region with densities $10^{14} < \nH < 10^{16} \ \percc$.
Table \ref{tab:PS} lists the disc mass $M_{\rm disc}$ and the mass accretion rate
$\dot{M}_{\rm disc}$.
$T_{\rm disc}$ is the effective temperature of the accreted gas onto the disc calculated
as $\dot{M}_{\rm disc} = \cs^3 / G \propto T_{\rm disc}^{3/2}$ \citep{Shu77}.
In Fig. \ref{fig:tmtot}, the solid curves show the total mass of the disc and protostars
(in the regions with densities $> 10^{14} \ \percc$), and the dashed curves show the total
mass of protostars with densities $> 10^{16} \ \percc$.
The difference of the solid curves and the dashed curves
indicates the disc mass as a function of time. 
In each halo, $\dot{M}_{\rm disc}$ decreases roughly monotonically with increasing metallicity.
The temperature $T_{\rm disc}$ represents the temperature at $\nH \sim 10^{14} \ \percc$ and
decreases owing largely to dust cooling (Fig. \ref{fig:nT}).
The accretion rate decreases with decreasing temperature, which is a consequence of the quasi-static condition set by the balance between gravitational force and
pressure gradient.
The relationship between $\dot{M}_{\rm disc}$ and metallicity is opposite for {\tt HBZ0} and {\tt HBZ-6}.
For {\tt HBZ0}, two nearly equal-mass protostars merge just after
PS1 formation ($t_* = 10.2$ yr).
The protostars are disrupted, and a fraction of their material
is stripped off into the ISM.
Since the ambient gas acquires angular momentum,
the accretion rate is suppressed
(see Appendix \ref{sec:nofrag} for detail).
The accretion rate for Halo B is generally larger than Halo A because the collapse timescale of
the cloud itself is shorter.

Figs. \ref{fig:rn_H0522} and \ref{fig:rn_H0611} show the profiles of density,
temperature, and radial velocity as a function of distance from the center of mass
of the most massive, primary protostar.
To see the three-dimensional properties of the accretion flow, we show the profiles in opposite directions
(solid and dashed curves) along the two axes parallel and perpendicular to the
angular momentum vector $\vb{J}$ (blue and orange curves, respectivley). 
We calculate total angular momentum as $\vb{J} = \sum m_i (\vb{r}_i \times \vb{v}_i)$,
where $\vb{r}_i$ is the displacement from the center of mass, and $\vb{v}_i$ is
the relative velocity to the mass-weighted mean velocity 
among cells with densities $> 10^{14} \ \percc$.
Outside the protostar ($\gtrsim 1$ au), the density suddenly declines down to 
$\sim 10^{13} \ \percc$ on the disc plane (orange curves)
and $\sim 10^{12} \ \percc$ along the rotational axis (blue curves).
The orange curves show the density plateau in the disc ($\sim 1$--$30$ au).
Along the rotational axis (blue curves), rapid gas accretion occurs with infall velocity
$| v_{\rm rad} | > 10$ km/s because the centrifugal force barrier does not work.
Since the infall velocity exceeds the local sound speed, accretion shocks form, and
the temperature increases to above $10^4$ K.
For {\tt HBZ-6}, the disc size is smaller than the other models,
and the gas is accreted onto the protostar
in a nearly spherical manner as we discuss in detail in Appendix \ref{sec:nofrag}.
Because of rapid gas accretion, the gas temperature
around the protostar rises up to $\sim 10^5$ K
due to compressional heating.

\subsubsection{Fragmentation of discs}

We find that DF occurs, and multiple protostellar systems form
in all our runs but for {\tt HBZ-6}
(Figs. \ref{fig:snapshots_acc_H0522_nH} and \ref{fig:snapshots_acc_H0611_nH}).
We identify candidate protostars by employing a friends-of-friends (FOF) algorithm
with the threshold density $n_{\rm H,th} = 10^{16} \ \percc$ 
and a linking length $4 \Delta x_i$.
We count the candidates with mass more than $0.003 \ \Msun$ ($\sim 3~\text{Jupiter masses}$)
as protostars.
Hereafter, the protostar born in the $n$'th order is dubbed as PS$n$ in each run.

Table \ref{tab:PS} shows that
as many as 26 protostars form until the time $t_{*, {\rm fin}}$ when we terminate the simulations.
Most of the protostars are destroyed, and only several protostars
remain in the accretion discs at $t_{*, {\rm fin}}$. 
The mass evolution of the protostellar systems shows the oligarchic growth.
This is consistent with the results of earlier works 
investigating the evolution of primordial protostellar discs \citep{Clark11, Stacy12, Greif12, Susa19}.
\citet{Susa19} derived an empirical relationship between the number of protostars and
elapsed time $t_*$ as
\begin{equation}
N_* = 3 \left( \frac{\tau _*}{1 \ {\rm yr}} \right)^{0.3}
\end{equation}
from earlier studies,
where $\tau _* = t_* (n_{\rm H, th} / 10^{19} \ \percc)^{1/2}$ is the time
scaled by the threshold density above which the gas is assumed to be adiabatic.
With $n_{\rm H,th} = 10^{16} \ \percc$ and $t_* = 100$--$400$ yr in this study,
the predicted number of fragments
is 4--6, which is consistent with our simulation results in both the metal-free and
metal-poor cases.

Most protostars form in dense spiral arms around the most massive protostar.
The protostars lose their angular momentum from the non-axisymmetric structure.
Then, they migrate toward PS1 in the manner of Type I migration as reported in
\citet{Greif12} and \citet{Hosokawa16}.
The migration time scale is comparable to the free-fall time
$\tff = 4.5$ yr in the region with density $\nH = 10^{14} \ \percc$.
The life time of the protostars which merge with PS1 in our simulations 
is found to be typically $\sim 1$--$10$ yr (see Appendix \ref{appendix}).
Since this is comparable to the free-fall timescale, we conclude that
the protostars undergo Type I migration \citep[also see][]{Chon19, Liao19}
and merge with the central protostar.


\subsubsection{Mass evolution of protostars}

The dashed curves of Fig. \ref{fig:tmtot} show the evolution of the total mass of 
protostars $M_{*,{\rm tot}}$, 
which monotonically increases through gas accretion.
Table \ref{tab:PS} shows $M_{*,{\rm tot}}$ at the time $t_{*, {\rm fin}}$ and
the average mass accretion rate $\dot{M}_*$ onto the protostars from $t_* = 0$ to $t_{*, {\rm fin}}$.
For Halo A,
the total mass becomes $M_{*,{\rm tot}} = 0.521$--$2.60 \ \Msun$ at 200--400 yr after PS1 formation.
The values correspond to the mass accretion rates $\dot{M}_* = 1.30\E{-3}$--$1.30\E{-2} \ \Msun / {\rm yr}$.
The accretion rate corresponds to the effective temperature 
$T_* = 278$--$1287$ K
calculated with a simple scaling $\dot{M}_* = \cs ^3 / G \propto T_*^{3/2}$. 
The mass accretion rate decreases with the increasing metallicity
because molecular and dust cooling reduces the gas temperature around the protostars.
For Halo B, 
the total mass is $M_{*,{\rm tot}} = 1.42$--$5.62 \ \Msun$ at 100--200 yr after PS1 formation,
corresponding to the mass accretion rates
$\dot{M}_* = 7.11\E{-3}$--$5.62\E{-2} \ \Msunyr$ and similarly evaluated temperatures $T_* = 862$--$3416$ K.
The mass accretion rate for Halo A is lower than that for Halo B with a fixed metallicity
because
the gravitational potential of MHs, $\propto GM_{\rm vir} / R_{\rm vir}$,
is shallower for Halo A ($1.2\E{11} \ {\rm erg \ g^{-1}}$)
than for Halo B ($2.7\E{11} \ {\rm erg \ g^{-1}}$).
Also, the gas in Halo A rotates more rapidly than Halo B and thus the
centrifugal force effectively prevents the collapse and accretion.
The rotation velocity of a cloud with mass $M$ and density $\rho$ is parametrized as 
\begin{equation}
\beta = \frac{E_{\rm rot}}{E_{\rm grav}} =
\frac{25}{12} \left( \frac{4\pi}{3} \right) ^{1/3} \frac{J^2 }{ G \rho ^{-1/3} M^{10/3}}.
\label{eq:beta}
\end{equation}
In this equation, $E_{\rm rot} \sim (\Omega R)^2 \sim (JR/I)^2$ is the rotational energy,
where $\Omega$, $I \sim MR^2 \sim M^{5/3}\rho^{-2/3}$ and $J=I\Omega$ are angular velocity,
momentum of inertia and total angular momentum, respectively,
and $E_{\rm grav} \sim GM/R$ is the gravitational energy.
The parameter $\beta = 0.033$ of Halo A
is larger than $\beta= 0.005$ of Halo B by a factor of seven.

Figs. \ref{fig:tm_H0522} and \ref{fig:tm_H0611} show the evolution of mass, accretion rate, distance
from the center of mass and the angular momentum of each protostar for Halo A and B, respectively.
Every protostar accretes the mass at the nearly constant rate around the average accretion
rate shown in Table \ref{tab:PS} (black dashed lines).
All protostars are bound in the gravitational potential of the disc, and remain within $\lesssim 100$ au
from the center of mass of the system.
Several researchers reported that some protostars are ejected from a disc through $N$-body
interactions \citep{Clark11, Greif12}.
We do not see this slingshot effect in our simulations for a few 100 yr of the disc evolution.
The protostars also accrete angular momentum, and the pamameter $j^2$ 
\begin{equation}
j^2 = \frac{J^2 }{ 4\pi G \rho _* ^{-1/3} M_* ^{10/3} }
\label{eq:j2}
\end{equation}
of protostars increases initially.
The angular momentum of PS1 suddenly decreases when it is torn apart into two fragments
by the centrifugal force (Section \ref{sec:protostar_formation}).
After the spiral arms develop in the discs, $j^2$ evolves nearly constantly due to the transport of
angular momentum through the non-axisymmetric structure.

\subsection{Formation/destruction processes of protostars}
\label{sec:protostar_formation_destruction}

We find that protostars form explosively within the first $\sim 10$ yr
but most of them are destroyed for $\sim 100$ yr.
In this section, we discuss the formation/destruction processes of
protostars on accretion discs and estimate the criterion for each process.
We show some examples of the processes in Appendix \ref{appendix}.

\subsubsection{Formation processes of protostars}
\label{sec:protostar_formation}

The formation of protostars is classified into the following three processes:

\vspace{0.5cm}
\noindent{{\bf (i) Gravitational contraction}}

In all runs, the primary protostar, PS1, forms through gravitational contraction and
grows mostly through gas accretion from filaments before accretion discs appear.
PS1 continues to be the most massive protostar until the simulation is terminated
(see red curves in Figs. \ref{fig:tm_H0522} and \ref{fig:tm_H0611})
in all runs but for {\tt HBZ0} where PS2 eventually becomes the most massive one.
For very low metallicities $\leq10^{-6} \ \Zsun$,
several subsequent protostars also form through gravitational contraction.
H$_2$ cooling induces the deformation of cloud cores, and dense filaments form, where
the secondary protostars form due to
gravitational instabilities within $t_* \sim 15$ yr.

\vspace{0.5cm}
\noindent{{\bf (ii) Break-up of rapidly rotating protostars}}

In eight out of ten runs (all but for {\tt HAZ-3} and {\tt HBZ-6}),
after PS1 sufficiently accretes mass and angular momentum, it breaks up
into two protostars when the centrifugal force overcomes the
gravitational force.
Until we terminate the simulation,
protostars forming through break-up mostly survive because
their initial orbital angular velocity is comparable to the escape velocity
of the primary protostar.
The protostars forming in this way become eventually as the second most massive protostars.

The criterion for break-up can be expressed by the critical angular momentum.
As in \citet{Eriguchi82}, we introduce a dimensionless parameter $j^2$ (Eq. \ref{eq:j2}),
which is the same order as $\beta$ (Eq. \ref{eq:beta}).
The lower panels of Figs. \ref{fig:tm_H0522} and \ref{fig:tm_H0611} show the evolution of
$j^2$.
Until the break-up occurs, $j^2$ of PS1 constantly increases along with mass accretion.
We find that the break-up criterion is
\begin{equation}
j^2 > j^2_{\rm cr} = 0.02
\end{equation}
with the critical value $j^2_{\rm cr}$ that we empirically determine.
The grey shades in Figs. \ref{fig:tm_H0522} and \ref{fig:tm_H0611}
indicate the region with $j^2 > j^2_{\rm cr}$. 
In our simulations, $j^2$ of PS1 reaches around $j^2_{\rm cr}$ and breaks up
in all cases
except for {\tt HAZ-3} and {\tt HBZ-6}, where no protostars form through the break-up
process.

We estimate the lower-limit of the initial $\beta$
of a cloud with a density $n_{\rm H, cl}$, 
above which the break-up of protostars occurs after the
gas density increases up to a density $n_{\rm H, *}$.
Assuming the conservation of angular momentum 
in the region containing a mass $M_*$,
we can derive the critical value of $\beta$ as
\begin{equation}
\beta _{\rm cr} = 4\E{-5}
\left( \frac{n_{\rm H,cl}}{10^{3} \ \percc} \right) ^{1/3}
\left( \frac{n_{\rm H,*}}{10^{16} \ \percc} \right) ^{-1/3}
\left( \frac{j^2_{\rm cr} }{0.02} \right).
\end{equation}
The range of $\beta$ of MHs are 0.1--1 \citep{Hirano14}, which suggests that
protostars forming in MHs can generally fragment through break-up.
We note that, in actual protostellar discs,
angular momentum is transported through non-axisymmetric 
structures, and then larger $\beta_{\rm cr}$ would be necessary for break-up
so that the protostars can obtain sufficient rotational energy larger than $j^2_{\rm cr}$.

\vspace{0.5cm}
\noindent{{\bf (iii) Interaction of spiral arms}}

There is yet another mode of protostar formation through the interaction of spiral arms in an accretion disc.
The gas at the colliding interface between two spiral arms is compressed and
becomes gravitationally unstable.
\citet{Takahashi16} and \citet{Inoue18} found that a spiral arm is unstable to fragment
when its minimum Toomre $Q$ parameter, $Q = \cs \kappa / \pi G \Sigma$, is below 0.6, 
where $\kappa$ is the epicyclic frequency, and $\Sigma$ is the gas surface density.
In this work, we also confirm that fragmentation occurs when
\begin{equation}
Q < 0.6
\label{eq:Q}
\end{equation}
after spiral arms interact as discussed in Appendix \ref{appendix}.

\subsubsection{Destruction processes of protostars}
\label{sec:protostar_destruction}

Although at most 26 protostars form in an accretion disc in the first few hundred years,
only 2--5 protostars survive after the destruction processes of protostars.
This suggests that the destruction process of protostars plays an important role in
the evolution of protostellar systems.
The destruction of protostars is classified into the following three processes:

\vspace{0.5cm}
\noindent{{\bf (i) Dissolution into ISM}}

Some protostars dissolve into ISM.
They are born with very small masses $\lesssim 0.01 \ \Msun$.
These protostars lose their mass because they are inefficiently self-gravitated.
We should note that some protostars just become not to
satisfy our mass criterion for a protostar $M_* > 0.003 \ \Msun$.
We do not follow further hydrodynamic evolution of such protostars.

\vspace{0.5cm}
\noindent{{\bf (ii) Mergers with another member protostar}}

Some protostars are destroyed through the interaction with another protostar.
The interaction can be divided into two categories: mergers (MER) and tidal disruption events (TDEs).
We define the criterion of MER and TDE by comparing the impact parameter
of an approaching protostar and the radius of a target protostar.
For a protostar PS$i$ approaching a protostar PS$j$, we calculate the impact parameter $b$
from the position and velocity of PS$i$ relative to PS$j$ from the snapshot 
taken just before the collision.
Then, we calculate the radius $a_j$ along the major axis of PS$j$ by approximating its shape to be
an ellipsoid.
We define the criteria for MER and TDE as 
\begin{eqnarray}
  \begin{cases}
    b < a_j & \text{MER (head-on collision)}, \\
    b > a_j & \text{TDE (off-set collision)}.
  \end{cases}
\end{eqnarray}

It has been believed that a TDE occurs when a protostar approaches another protostar within the tidal radius $r_{\rm tid} = a_j (m_j / m_i)^{1/3}$, where $m_i$ and $m_j$ are the mass
of PS$i$ and PS$j$, respectively \citep{Rees1988}.
However, we find that TDEs occur even when $b > r_{\rm tid}$ (Appendix \ref{appendix}).
Since the masses of the colliding protostars are comparable ($m_i / m_j \sim 1$),
both protostars deform with each other's tidal force.
The cross-section for the interaction can become effectively larger than in the case 
with a large mass ratio ($m_j / m_i \gg 1$) often assumed by the classical theory of TDEs, 
where only a smaller protostar deforms.

\vspace{0.5cm}
\noindent{{\bf (iii) Tidal disruption events}}

Some protostars are destroyed through TDEs when they collide with each other
in an off-set manner.
In some TDEs, a new protostar forms in dense tidal tails
(Appendix \ref{appendix}).

\begin{figure*}
\includegraphics[width=18cm]{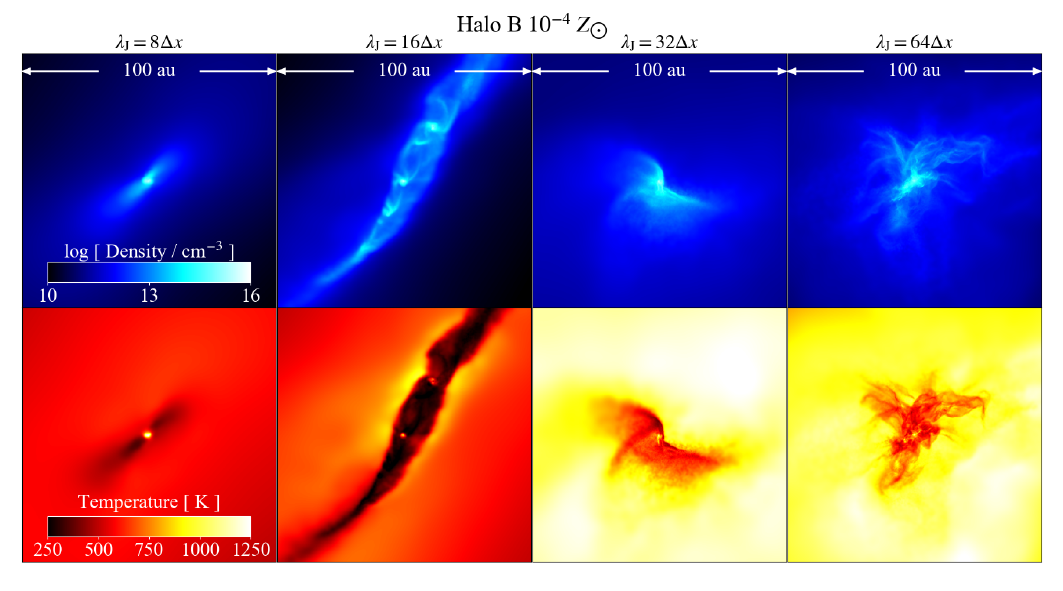}
\caption{Density-weighted density projections of Halo B with a metallicity $Z=10^{-4} \ \Zsun$
when the maximum density reaches $\nHmax = 10^{16} \ \percc$.
We show the results for the resolution criteria of $\ljeans = 8$, $16$, $32$, and $64 \Delta x$ from left to right.}
\label{fig:snapshots_res_H0611_Z-4}
\end{figure*}

\section{Discussion}
\label{sec:discussion}

\subsection{Final mass of metal-free/metal-poor stars}

In this work, we follow the evolution of accretion discs for $100$--$400$ yr,
and find that the protostars grow up to $\sim 0.1$--$1 \ \Msun$.
Fig. \ref{fig:tmtot} shows the protostellar systems still
accrete the gas and grow in mass.
In principle, it is necessary to follow
the disc and protostellar evolution for over $\sim 10^5$ yr until the protostars reach ZAMS, 
in order to determine the mass and the number of stars forming in the system.
Further fragmentation and gas accretion onto protostars will be halted by the 
radiative feedback when the primary protostar grows up to $7 \ \Msun$ and
starts emitting ultraviolet photons \citep{Hosokawa16, Fukushima20, Sugimura20}.
If (a part of) protostars remain low-mass ($< 1 \ \Msun$), they will be observed as
extremely metal-poor stars.
If they can grow up to $\sim 10 \ \Msun$ and if the number of protostars does not
significantly increase, massive metal-poor binaries or clusters will form,
which can explain the
origin of gravitational waves recently detected by LIGO/VIRGO
because the mass-loss rate of metal-free and metal-poor stars 
is small in their main sequence, and most of their mass can contract into black holes \citep{Abbott15}.
To explicitly see the final fate of the protostellar system, we require
another numerical strategy to follow the longer-term evolution of
accretion discs, such as the sink particle technique.
This technique requires some modeling to describe the interaction of sinks.
This work can supply realistic models of the mergers and TDEs between sinks.

\subsection{Initial metal/dust models}
The accretion rate decreases with the increasing dust cooling rate, which depends on
the abundance and size distribution of dust grains.
In this work we fix the initial metal and dust properties for a Pop III SN model
with a progenitor mass $\MPopIII = 30 \ \Msun$ and an ambient gas density $\namb = 1 \ \percc$.
With a simple semi-analytic model, \citet{Chiaki15} found that the thermal evolution of
clouds depends on progenitor models.
For $\MPopIII = 13 \ \Msun$, the growth of carbon grains also enhances the dust cooling rate
because the elemental abundance ratio of carbon to oxygen is larger than for more massive
progenitor mass.
With increasing $\namb$, depletion efficiency of metals onto grains decreases because of
the increasing rate of sputtering by reverse shocks.
The critical metallicity above which dust cooling can operate increases 
for $\namb = 10 \ \percc$ than in the case without grain destruction by an order of
magnitude.
To statistically investigate the mass spectrum of Pop II stars, we need to 
consider the realistic frequency of $\MPopIII$ and $\namb$.

This work can apply to the formation of metal-poor stars with peculiar
elemental abundance patterns, such as CEMP stars \citep{Beers05}.
Also for clouds with different elemental abundances,
we can discuss the condition for CF/DF on the analogy of this work.
In our recent work \citep{Chiaki20}, we followed the collapse of
gas clouds with C-enhanced elemental abundance ratios
until the formation of the primary protostar.
We employed a ``faint SN'' model, where Fe-rich innermost layers fall back into a central
SN remnant, and relatively C-enhanced gas is ejected \citep{Umeda03}.
The enrichment from faint SNe is one of the scenarios that can explain the formation of CEMP stars.
We found that the cloud core fragments due to rapid gas cooling
from amorphous carbon grains for the lowest progenitor mass ($\MPopIII = 13 \ \Msun$).
For more massive progenitor mass ($\MPopIII = 50$ and $80 \ \Msun$), 
the clouds collapse stably even with efficient dust cooling because of H$_2$
formation heating.
Interestingly, for $\MPopIII = 50 \ \Msun$, we found that spiral arms develop around the primary
protostar, where the Toomre $Q$ parameter is below the critical value 0.6.
This suggests that DF would occur if we follow the longer-term evolution
of the arms.

In this work, we have assumed the elemental abundance of a single Pop III SN model.
Multiple Pop III and Pop II progenitors can contribute to metal enrichment
of a halo.
\citet{Hartwig18} shows that C-normal stars with ${\rm [Mg/C]} > -1$
are likely to form in clouds enriched by multiple progenitors.
To compare with simulation results with the large samples of observed EMP stars,
we are required to consider multiply enriched gas clouds.


\begin{figure}
\includegraphics[width=\columnwidth]{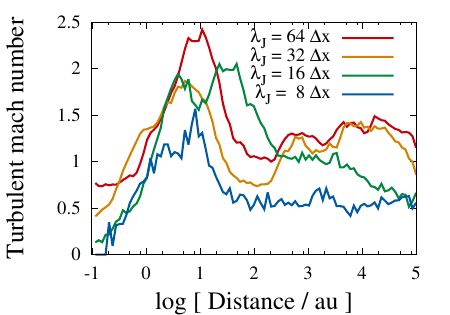}
\caption{Turbulent mach number as a function of 
distance from the density maximum for {\tt HBZ-4}
when the maximum density reaches $\nHmax = 10^{16} \ \percc$.
The red, orange, green and blue curves for $\ljeans = 8$, $16$, $32$, and $64 \Delta x$, respectively.}
\label{fig:rMturb}
\end{figure}

\subsection{Numerical resolution}
\label{sec:resolution}

We have imposed the resolution criterion where the local Jeans length $\ljeans$ is resolved by 32 cells.
It is smaller than other simulations of Pop III/II star formation \citep{Greif12, Smith15, Chiaki19}.
In this section, we see whether the cloud morphology, which significantly affects the fragmentation
property, converges or not with our resolution criterion.
We also aim to compare the present work with our previous work \citepalias{Chiaki16}, where
we used the same initial conditions and chemistry/cooling models but imposed milder
resolution criterion ($\ljeans = 10 \Delta x$).

Fig. \ref{fig:snapshots_res_H0611_Z-4} shows the projections of density and temperature of the
clouds for {\tt HBZ-4} with criteria $\ljeans = 8$, $16$, $32$ and $64 \Delta x$.
The shape of the clouds depends on resolution.
With higher resolution, turbulence is resolved at smaller spatial scales.
For the lowest resolution, $\ljeans = 8 \Delta x$, the cloud collapses in an almost spherical manner.
For intermediate resolutions, $\ljeans = 16$ and $32 \Delta x$, the cloud shape
deviates from the spherical form.
For the highest resolution, $\ljeans = 64 \Delta x$, small filaments with a size of $\sim 10$ au
appear around the central dense protostar.

Fig. \ref{fig:rMturb} shows the radial profile of turbulent mach number
$\Mturb = |\vb{v}_{\rm turb}| / \cs$.
Following \citet{Higashi21},
we estimate the turbulent velocity $\vb{v}_{\rm turb}$ by taking the mass-weighted root-mean-square of
\begin{equation}
\vb{v}_{{\rm turb},i} = \vb{v}_i - \vb{v}_{\rm mean} - \vb{v}_{\rm rad} - \vb{v}_{\rm rot}
\end{equation}
of a cell $i$
to subtract the bulk motion including mean velocity $\vb{v}_{\rm mean}$,
radial velocity $\vb{v}_{\rm rad}$ and rotational velocity $\vb{v}_{\rm rot}$  
in each of 100 logarithmic bins from a radius 0.1 au to 1 pc.
The amplitude of turbulence is monotonically larger for higher resolution at scales $\sim 10$ au,
because the numerical diffusion is smaller.

Interestingly, for $\ljeans = 16 \Delta x$,
a dense filamentary structure appears at a scale $\sim 50$ au 
(Fig. \ref{fig:snapshots_res_H0611_Z-4}).
This cloud morphology is consistent with the result of \citetalias{Chiaki16} 
(see their figure 5).
At this scale, turbulence is amplified up to $\Mturb = 2$
(the green curve in Fig. \ref{fig:rMturb}).
The scale may be a ``resonant'' scale of turbulent amplification 
for this resolution.
Fig. \ref{fig:snapshots_res_H0611_Z-4} shows that the filament is defined 
by accretion shocks with
a density $\sim 10^{12} \percc$ and a temperature $\sim 1000$ K,
corresponding to the Jeans length of $70$ au. 
This indicates that the turbulence is powered by the accretion shocks.

Although the cloud morphology still does not converge at the scales $\lesssim 10$ au 
for $\ljeans = 32 \Delta$, we use this criterion to follow the long-term evolution of accretion discs
with the present computational power.

\subsection{Effect of the chimistry model on break-up}

We find that several protostars form through the break-up of the primary protostar, but
the density structure in protostars may affect their hydrodynamic evolution.
In this study, we impose the stiff EOS, and the density is almost uniform at densities above
$n_{\rm H, th} = 10^{16} \ \percc$ because of the intense pressure 
(Figs. \ref{fig:rn_H0522} and \ref{fig:rn_H0611}).
In reality, the specific heat ratio is $\gamma \sim 1.1$ at densities $10^{16} < \nH < 10^{21} \ \percc$
with gas cooling accompanied by the destruction of H$_2$ formation while 
$\gamma \sim 1.4$ at $\nH > 10^{21} \ \percc$ \citep{Omukai00}.
Since the density is proportional to $\propto r^{-2/(2-\gamma)}$, a less-dense envelope should extend
around the core, which might suppress the break-up.
However, \citet{Boss79} found that a rotating isothermal cloud deforms into an array-like shape and fragment into two blobs.
\citet{Greif12} reported that break-up occurs in primordial clouds with a realistic chemistry
model.
We argue that the break-up still occurs with a realistic chemistry model
although its frequency might be suppressed by the density structure in cloud cores.
We will investigate the effect of the critical density on the break-up process in the forthcoming paper.

\section{Summary and conclusion}
\label{sec:conclusion}

It has been conventionally postulated that the typical mass scale becomes smaller from Pop III to Pop II
stars because additional cooling from dust thermal emission induces CF
as metallicity increases.
In this study, we find that CF does not occur in most runs
because of rapid gas heating associated with H$_2$ molecular formation.
Although CF occurs in several runs, fragments are rapidly accreted onto the primary protostar.
Instead, the dominant mode of fragmentation is DF
regardless of metallicities.
At most 26 protostars form through gas accretion or break-up, but
most of them merge with each other or destroyed through TDEs.
At the endpoint of our simulations, only several protostars remain in the systems.
We also find that oligarchic evolution of the protostars occurs;
low-mass protostars orbit around the central massive star or around a massive binary.

The mass distribution and multiplicity of the stars are determined
only after the gas accretion and fragmentation are halted.
This occurs when the mass of the most massive protostar(s) 
exceeds $\sim 7 \ \Msun$, and its radiative feedback is effective \citep{Hosokawa16}.
Whether low-mass or high-mass stars form depends on the strength and the timing 
of the radiative feedback.
In future works, we will employ advanced numerical techniques, such  
as sink particles \citep[see][]{Fukushima20}, and will follow the long-term evolution
of low-metallicity protostars to uncover the entire picture of metal-free/metal-poor
star formation.

\section*{ACKNOWLEDGMENTS}

We thank T. Nozawa, who kindly gave us his SN model.
The initial conditions in the simulations are provided by S. Hirano. 
We thank the fruitful discussion with K. Omukai, H. Susa and S. Higashi.
GC is supported by Overseas Research
Fellowships of the Japan Society for the Promotion of Science (JSPS)
for Young Scientists.
NY acknowledges support by SPPEXA through JST CREST JPMHCR1414.
The numerical simulations and analyses in this work are carried out on
XC40 in Yukawa Institute of Theoretical Physics (Kyoto University), and
COMET in SDSC.

\section*{Data availability}

The versions of {\sc grackle}
used in this work is available at 
\url{https://github.com/genchiaki/grackle/tree/metal-dust-radiation}.
The data underlying this article will be shared on reasonable request
to the authors.


\label{lastpage}

\appendix
\section{Evolution of protostellar systems}
\label{appendix}

\begin{table*}
\begin{minipage}{15cm}
\caption{Properties of protostars for Halo A}
\label{tab:PSs_HA}
\begin{tabular}{cccccccccccccc}
\hline
 $Z$       & $i$ & $t_{\rm form}$ & form  &parent & $t_{\rm dest}$ & dest. & $j$  & $b$  &  $a_j$    & $m_j/m_i$& $r_{\rm tid}$ & $t_{\rm life}$ &     $M_*$   \\
 $[\Zsun]$ &     & [yr]           &       &       & [yr]           &       &      & [au] &  [au]     &          & [au]          & [yr]           & [$\Msun$]   \\
\hline \hline                                                                                       
 $10^{-3}$ &   1 &            2.7 &    GC &   --- &            --- &   --- & ---  &  --- &       --- &       ---&           --- &            --- &      0.471  \\
           &   2 &          368.7 &   INT &   --- &            --- &   --- & ---  &  --- &       --- &       ---&           --- &            --- &      0.025  \\
           &   3 &          383.3 &   INT &   --- &            --- &   --- & ---  &  --- &       --- &       ---&           --- &            --- &      0.026  \\
\hline                                                                                                                                   
 $10^{-4}$ &   1 &            1.5 &    GC &   --- &            --- &   --- & ---  &  --- &       --- &       ---&           --- &            --- &      0.328  \\
           &   2 &          114.3 &    BU &    1  &            --- &   --- & ---  &  --- &       --- &       ---&           --- &            --- &      0.097  \\
           &   3 &          128.7 &   INT &   --- &            --- &   --- & ---  &  --- &       --- &       ---&           --- &            --- &      0.142  \\
\hline                                                                                                                                   
 $10^{-5}$ &   1 &            0.9 &    GC &   --- &            --- &   --- & ---  &  --- &       --- &       ---&           --- &            --- &      0.584  \\
           &   2 &           35.9 &   INT &   --- &           39.1 &   MER &   1  & 3.72 &      4.22 &     53.07&         15.87 &            3.2 & (    0.005) \\
           &   3 &           59.7 &   INT &   --- &           92.3 &   MER &   4  & 0.13 &      1.05 &      3.39&          1.57 &           32.6 & (    0.037) \\
           &   4 &           60.9 &    BU &    1  &          137.1 &   TDE &   1  & 3.82 &      2.03 &      1.44&          2.30 &           76.2 & (    0.283) \\
           &   5 &           68.7 &   INT &   --- &           80.9 &   TDE &   4  & 1.16 &      0.74 &      3.73&          1.15 &           12.2 & (    0.020) \\
           &   6 &           71.7 &   INT &   --- &          145.7 &   TDE &   8  & 3.03 &      1.51 &      1.63&          1.78 &           74.0 & (    0.110) \\
           &   7 &           75.3 &   INT &   --- &            --- &   --- & ---  &  --- &       --- &       ---&           --- &            --- &      0.140  \\
           &   8 &          141.5 &    BU &    1  &            --- &   --- & ---  &  --- &       --- &       ---&           --- &            --- &      0.309  \\
\hline                                                                                                                                   
 $10^{-6}$ &   1 &            0.3 &    GC &   --- &            --- &   --- & ---  &  --- &       --- &       ---&           --- &            --- &      0.795  \\
           &   2 &           53.9 &    BU &    1  &            --- &   --- & ---  &  --- &       --- &       ---&           --- &            --- &      0.580  \\
           &   3 &           87.1 &   INT &   --- &           89.5 &   MER &   2  & 1.00 &      1.19 &      7.57&          2.33 &            2.4 & (    0.016) \\
           &   4 &           89.3 &   INT &   --- &           90.9 &   TDE &   2  & 1.59 &      1.48 &      8.29&          3.00 &            1.6 & (    0.018) \\
           &   5 &          114.9 &    BU &    1  &            --- &   --- & ---  &  --- &       --- &       ---&           --- &            --- &      0.032  \\
\hline                                                                                                                                   
 0         &   1 &            0.3 &    GC &   --- &            --- &   --- & ---  &  --- &       --- &       ---&           --- &            --- &      1.782  \\
           &   2 &            3.1 &    GC &   --- &            3.3 &   MER &   1  & 0.00 &      4.13 &     17.87&         10.79 &            0.2 & (    0.005) \\
           &   3 &            3.3 &    GC &   --- &            7.7 &   MER &   1  & 0.61 &      1.27 &      4.51&          2.10 &            4.4 & (    0.044) \\
           &   4 &            4.9 &    GC &   --- &            6.3 &   MER &   1  & 0.32 &      1.11 &      9.66&          2.37 &            1.4 & (    0.016) \\
           &   5 &            4.9 &    GC &   --- &           11.7 &   MER &   1  & 0.87 &      1.39 &      2.96&          1.99 &            6.8 & (    0.107) \\
           &   6 &            8.5 &    GC &   --- &           13.5 &   MER &   1  & 1.02 &      2.63 &     26.32&          7.83 &            5.0 & (    0.017) \\
           &   7 &           11.9 &    GC &   --- &           15.7 &   MER &   1  & 1.36 &      2.14 &     17.50&          5.55 &            3.8 & (    0.029) \\
           &   8 &           87.9 &    BU &    1  &            --- &   --- & ---  &  --- &       --- &       ---&           --- &            --- &      0.712  \\
           &   9 &          112.7 &   INT &   --- &          123.5 &   DIS & ---  & 5.26 &      4.01 &     20.30&         10.93 &           10.8 & (    0.078) \\
           &  10 &          117.1 &   INT &   --- &          163.1 &   TDE &   1  & 5.64 &      4.16 &     15.11&         10.29 &           46.0 & (    0.110) \\
           &  11 &          123.7 &    GC &   --- &          123.9 &   TDE &   1  & 0.49 &      4.11 &     20.36&         11.23 &            0.2 & (    0.077) \\
           &  12 &          137.1 &   INT &   --- &          154.1 &   MER &   8  & 0.22 &      1.05 &      2.56&          1.44 &           17.0 & (    0.111) \\
           &  13 &          168.5 &    BU &    1  &            --- &   --- & ---  &  --- &       --- &       ---&           --- &            --- &      0.103  \\
\hline
\end{tabular}
\medskip \\
Note --- 
(1) Metallicity $Z$.
(2) ID of a protostar $i$. 
(3) Formation time $t_{\rm form}$ of protostars.
(4) Formation paths of each protostar: 
gravitational contraction (GC),
break-up of a parent protostar (BU), and 
interaction of spiral arms (INT).
(5) ID of the parent protostar which breaks up.
(6) Destruction time $t_{\rm dest}$ of protostars.
(7) Destruction mechanism: 
merger (MER),
tidal disruption event (TDE), and
dissolution to the ambient gas (DIS).
(8) ID of counterpart $j$ with which the protostar $i$ interacts.
(9-12) Impact parameter $b$, 
radius of major-axis $r_j$, mass radio $m_j/m_i$, and tidal radius $r_{\rm tid}$ 
of the impactor $j$.
(13) Life time $t_{\rm life}$ of the protostar.
(14) protostellar mass $M_*$ when it is destroyed with parentheses or
when the simulation is terminated at $t_{*,{\rm fin}}$ without parentheses.
\end{minipage}
\end{table*}

\begin{table*}
\begin{minipage}{15cm}
\caption{Properties of protostars for Halo B}
\label{tab:PSs_HB}
\begin{tabular}{cccccccccccccc}
\hline
 $Z$       & $i$ & $t_{\rm form}$ & form  &parent & $t_{\rm dest}$ & dest. & $j$  & $b$  &  $a_j$    & $m_j/m_i$& $r_{\rm tid}$ & $t_{\rm life}$ &     $M_*$   \\
 $[\Zsun]$ &     & [yr]           &       &       & [yr]           &       &      & [au] &  [au]     &          & [au]          & [yr]           & [$\Msun$]   \\
\hline \hline                                                                                       
 $10^{-3}$ &   1 &            0.7 &    GC &   --- &            --- &   --- & ---  &  --- &       --- &       ---&           --- &            --- &      1.407  \\
           &   2 &          178.9 &    BU &    1  &            --- &   --- & ---  &  --- &       --- &       ---&           --- &            --- &      0.015  \\
\hline                                                                                                                                 
 $10^{-4}$ &   1 &            1.3 &    GC &   --- &            --- &   --- & ---  &  --- &       --- &       ---&           --- &            --- &      1.693  \\
           &   2 &           52.3 &   INT &   --- &           67.9 &   MER &   1  & 5.96 &      6.46 &     28.11&         19.63 &           15.6 & (    0.044) \\
           &   3 &           69.9 &    GC &   --- &           70.1 &   DIS & ---  & 0.00 &      5.20 &    280.10&         34.00 &            0.2 & (    0.005) \\
           &   4 &           73.5 &    BU &    1  &            --- &   --- & ---  &  --- &       --- &       ---&           --- &            --- &      0.018  \\
           &   5 &           78.5 &    BU &    1  &            --- &   --- & ---  &  --- &       --- &       ---&           --- &            --- &      0.156  \\
           &   6 &           87.5 &   INT &   --- &            --- &   --- & ---  &  --- &       --- &       ---&           --- &            --- &      0.012  \\
           &   7 &           90.1 &   INT &   --- &           95.1 &   TDE &   1  & 5.83 &      4.21 &    328.25&         29.06 &            5.0 & (    0.005) \\
           &   8 &           92.5 &   INT &   --- &           93.5 &   TDE &   1  & 5.85 &      3.74 &    287.93&         24.68 &            1.0 & (    0.005) \\
           &   9 &           94.5 &    BU &    1  &           94.7 &   MER &   1  & 0.84 &      4.11 &    420.29&         30.81 &            0.2 & (    0.004) \\
\hline                                                                                                                                 
 $10^{-5}$ &   1 &            0.5 &    GC &   --- &            --- &   --- & ---  &  --- &       --- &       ---&           --- &            --- &      1.076  \\
           &   2 &           42.5 &    BU &    1  &            --- &   --- & ---  &  --- &       --- &       ---&           --- &            --- &      0.393  \\
           &   3 &           42.7 &   INT &   --- &           46.9 &   TDE &   1  & 3.11 &      2.52 &     76.74&         10.71 &            4.2 & (    0.009) \\
           &   4 &           52.3 &   INT &   --- &           87.1 &   TDE &   1  & 3.15 &      3.07 &     21.49&          8.54 &           34.8 & (    0.045) \\
           &   5 &           54.5 &   INT &   --- &           67.9 &   TDE &   2  & 1.38 &      1.25 &     16.93&          3.20 &           13.4 & (    0.010) \\
           &   6 &           58.7 &   INT &   --- &           66.7 &   TDE &   2  & 1.35 &      1.13 &     10.96&          2.52 &            8.0 & (    0.015) \\
           &   7 &           69.3 &   INT &   --- &           83.7 &   TDE &   2  & 1.57 &      1.25 &     20.51&          3.41 &           14.4 & (    0.013) \\
           &   8 &           82.3 &   INT &   --- &           94.5 &   MER &   1  & 1.64 &      3.78 &     41.05&         13.03 &           12.2 & (    0.025) \\
           &   9 &           83.9 &   INT &   --- &            --- &   --- & ---  &  --- &       --- &       ---&           --- &            --- &      0.011  \\
           &  10 &           89.1 &    BU &    1  &           89.7 &   DIS & ---  & 4.44 &      3.14 &    317.30&         21.45 &            0.6 & (    0.003) \\
           &  11 &           93.3 &    GC &   --- &           94.5 &   DIS & ---  &11.55 &      3.78 &    330.92&         26.12 &            1.2 & (    0.003) \\
           &  12 &           94.7 &    GC &   --- &            --- &   --- & ---  &  --- &       --- &       ---&           --- &            --- &      0.005  \\
           &  13 &           99.1 &    BU &    1  &           99.5 &   DIS & ---  & 4.46 &      3.59 &    196.66&         20.86 &            0.4 & (    0.005) \\
\hline                                                                                                                                 
 $10^{-6}$ &   1 &            0.5 &    GC &   --- &            --- &   --- & ---  &  --- &       --- &       ---&           --- &            --- &      5.615  \\
           &   2 &            8.3 &    GC &   --- &           11.7 &   MER &   1  & 1.17 &      1.99 &      9.60&          4.23 &            3.4 & (    0.078) \\
           &   3 &            9.5 &    GC &   --- &           13.1 &   MER &   1  & 1.80 &      2.55 &     15.96&          6.43 &            3.6 & (    0.057) \\
           &   4 &           11.5 &    GC &   --- &           14.5 &   TDE &   1  & 2.62 &      2.24 &     25.98&          6.65 &            3.0 & (    0.039) \\
\hline                                                                                                                                 
 0         &   1 &            0.3 &    GC &   --- &           10.3 &   TDE &   2  & 3.70 &      2.06 &      1.15&          2.15 &           10.0 & (    0.365) \\
           &   2 &            4.3 &   INT &   --- &            --- &   --- & ---  &  --- &       --- &       ---&           --- &            --- &      2.113  \\
           &   3 &           32.1 &   INT &   --- &           35.5 &   TDE &   2  & 3.92 &      2.61 &     16.01&          6.57 &            3.4 & (    0.056) \\
           &   4 &           32.5 &   INT &   --- &           36.3 &   MER &   7  & 0.76 &      0.77 &      0.32&          1.25 &            3.8 & (    0.036) \\
           &   5 &           32.7 &   INT &   --- &           39.1 &   TDE &   2  & 3.42 &      2.97 &      5.89&          5.36 &            6.4 & (    0.167) \\
           &   6 &           33.1 &    BU &    5  &           33.3 &   MER &   5  & 1.06 &      2.18 &      3.29&          3.25 &            0.2 & (    0.005) \\
           &   7 &           35.9 &   INT &   --- &           36.3 &   MER &   4  & 0.76 &      0.86 &      3.08&          1.25 &            0.4 & (    0.012) \\
           &   8 &           37.3 &    BU &    5  &           40.9 &   MER &  13  & 0.66 &      0.67 &      1.49&          0.77 &            3.6 & (    0.024) \\
           &   9 &           37.5 &   INT &   --- &           49.3 &   TDE &  13  & 3.87 &      3.73 &      3.88&          5.86 &           11.8 & (    0.133) \\
           &  10 &           37.5 &   INT &   --- &           47.7 &   TDE &  13  & 4.31 &      2.16 &      1.16&          2.27 &           10.2 & (    0.203) \\
           &  11 &           38.1 &    BU &    2  &           47.1 &   MER &  10  & 2.05 &      2.51 &      3.24&          3.71 &            9.0 & (    0.045) \\
           &  12 &           38.3 &   INT &   --- &           45.5 &   TDE &   9  & 1.88 &      0.90 &      2.77&          1.26 &            7.2 & (    0.031) \\
           &  13 &           38.9 &   INT &   --- &           66.5 &   TDE &   2  & 5.71 &      2.80 &      1.71&          3.35 &           27.6 & (    0.855) \\
           &  14 &           40.5 &   INT &   --- &           45.3 &   TDE &  13  & 1.24 &      1.05 &      0.98&          1.03 &            4.8 & (    0.097) \\
           &  15 &           40.5 &   INT &   --- &           49.3 &   MER &   2  & 3.22 &      4.10 &    305.09&         27.59 &            8.8 & (    0.004) \\
           &  16 &           41.1 &    BU &    2  &           41.7 &   TDE &  15  & 5.21 &      1.26 &      7.53&          2.47 &            0.6 & (    0.003) \\
           &  17 &           42.1 &   INT &   --- &           43.3 &   TDE &  15  & 5.49 &      0.90 &     12.91&          2.12 &            1.2 & (    0.004) \\
           &  18 &           42.7 &    BU &   11  &           43.7 &   TDE &  10  & 1.92 &      1.50 &      3.48&          2.28 &            1.0 & (    0.023) \\
           &  19 &           43.9 &   INT &   --- &           49.9 &   MER &   2  & 3.48 &      4.10 &     48.15&         14.91 &            6.0 & (    0.027) \\
           &  20 &           47.5 &   INT &   --- &           59.9 &   TDE &   2  & 4.84 &      4.33 &     12.93&         10.16 &           12.4 & (    0.109) \\
           &  21 &           51.5 &   INT &   --- &            --- &   --- & ---  &  --- &       --- &       ---&           --- &            --- &      0.352  \\
           &  22 &           65.9 &   INT &   --- &            --- &   --- & ---  &  --- &       --- &       ---&           --- &            --- &      0.126  \\
           &  23 &           69.9 &   INT &   --- &            --- &   --- & ---  &  --- &       --- &       ---&           --- &            --- &      0.815  \\
           &  24 &           70.7 &    BU &    2  &            --- &   --- & ---  &  --- &       --- &       ---&           --- &            --- &      0.365  \\
           &  25 &           71.5 &   INT &   --- &           73.9 &   TDE &  22  & 1.50 &      0.94 &      2.54&          1.28 &            2.4 & (    0.022) \\
           &  26 &           75.3 &   INT &   --- &           82.5 &   MER &  23  & 1.02 &      1.56 &      3.44&          2.36 &            7.2 & (    0.121) \\
\hline
\end{tabular}
\medskip \\
\end{minipage}
\end{table*}

\begin{figure*}
 \includegraphics[width=18.0cm]{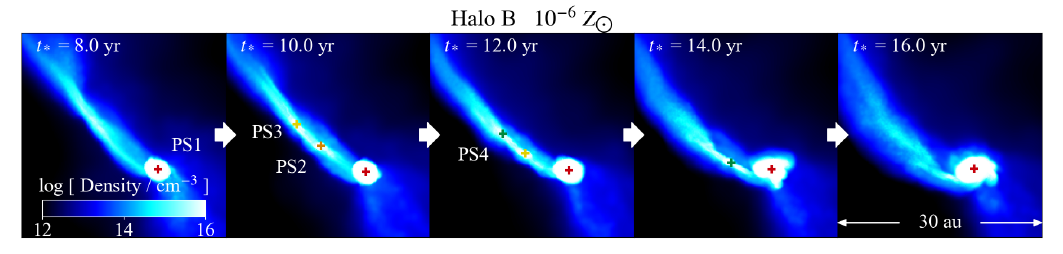}
\caption{
Projection of density in the accretion disc for {\tt HBZ-6}
from $t_* = 8$ to 16 yr, where
the gravitational contraction of the dense filament drives protostar formation.
}
\label{fig:snapshots_acc_acc}

 \includegraphics[width=18.0cm]{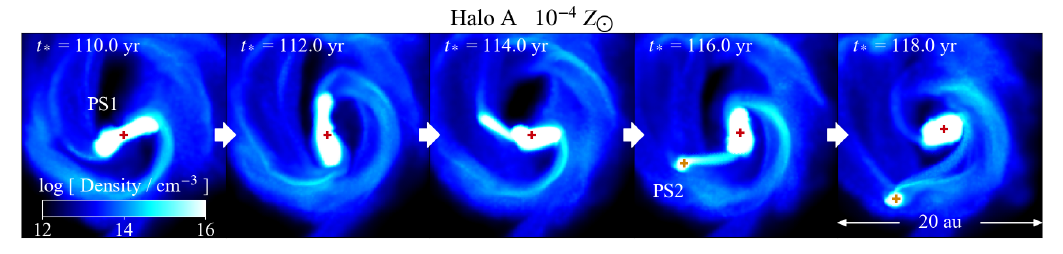}
\caption{
Projection of density in the accretion disc for {\tt HAZ4}
from $t_* = 110$ to 118 yr, where
the break-up of a rapidly rotating protostar PS1 generates a protostar PS2.
}
\label{fig:snapshots_acc_bu}
\end{figure*}

\begin{figure*}
 \includegraphics[width=18.0cm]{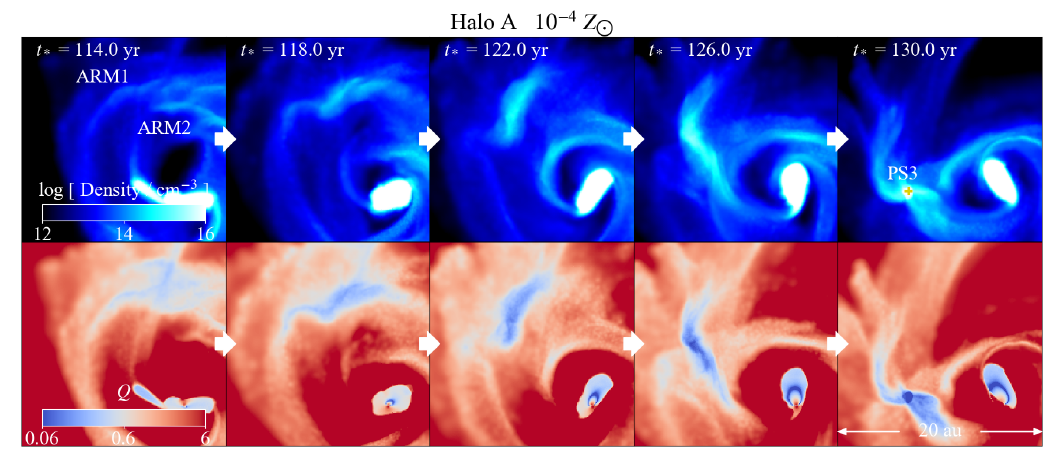}
\caption{
Projections of density (top panels) and Toomre $Q$ parameter (bottom panels)
in the accretion disc for {\tt HAZ-4} from $t_* = 114$ to 130 yr, where
the interaction between two spiral arms ARM1 and ARM2 drives formation of a protostar PS3.
}
\label{fig:snapshots_acc_int}
\end{figure*}

\begin{figure*}
 \includegraphics[width=18.0cm]{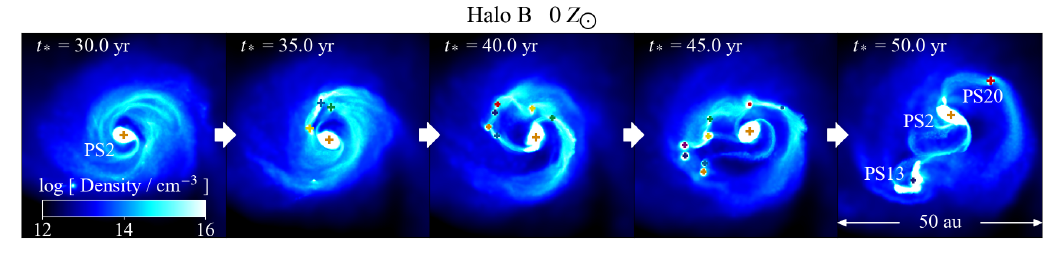}
\caption{
Projection of density in the accretion disc 
for {\tt HAZ0} from $t_* = 30$ to 50 yr, 
where $>20$ protostars rapidly forms.
}
\label{fig:snapshots_acc_H0611_Z-9}
\end{figure*}

\begin{figure*}
 \includegraphics[width=18.0cm]{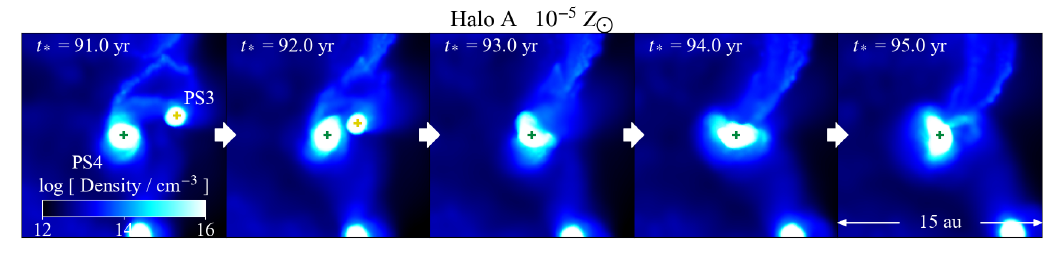}
\caption{
Projection of density in the accretion disc 
for {\tt HBZ-5} from $t_* = 91$ to 95 yr, 
where merger of protostars occurs.
}
\label{fig:snapshots_acc_mer}
\end{figure*}

\begin{figure*}
 \includegraphics[width=18.0cm]{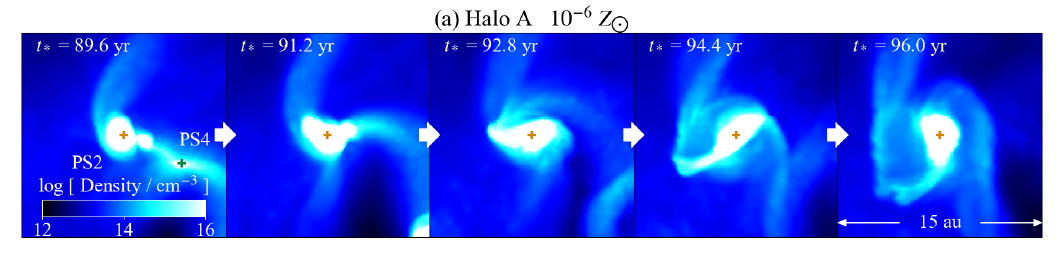}
 \includegraphics[width=18.0cm]{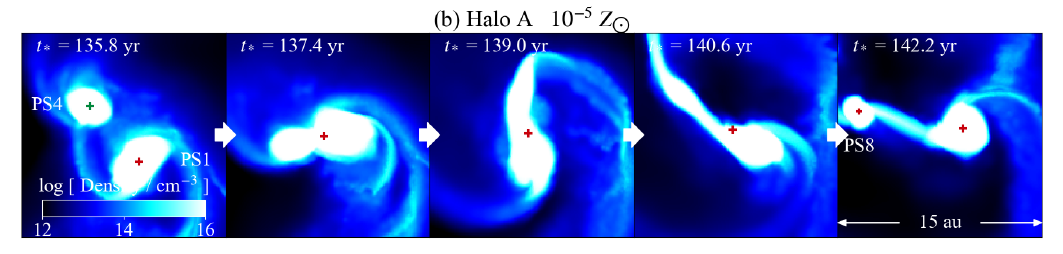}
\caption{
Projection of density in the accretion discs 
for (a) {\tt HAZ-6} and (b) {\tt HAZ-5}, where tidal disruption events (TDEs) occur.
}
\label{fig:snapshots_acc_tde}
\end{figure*}

\begin{figure*}
 \includegraphics[width=18.0cm]{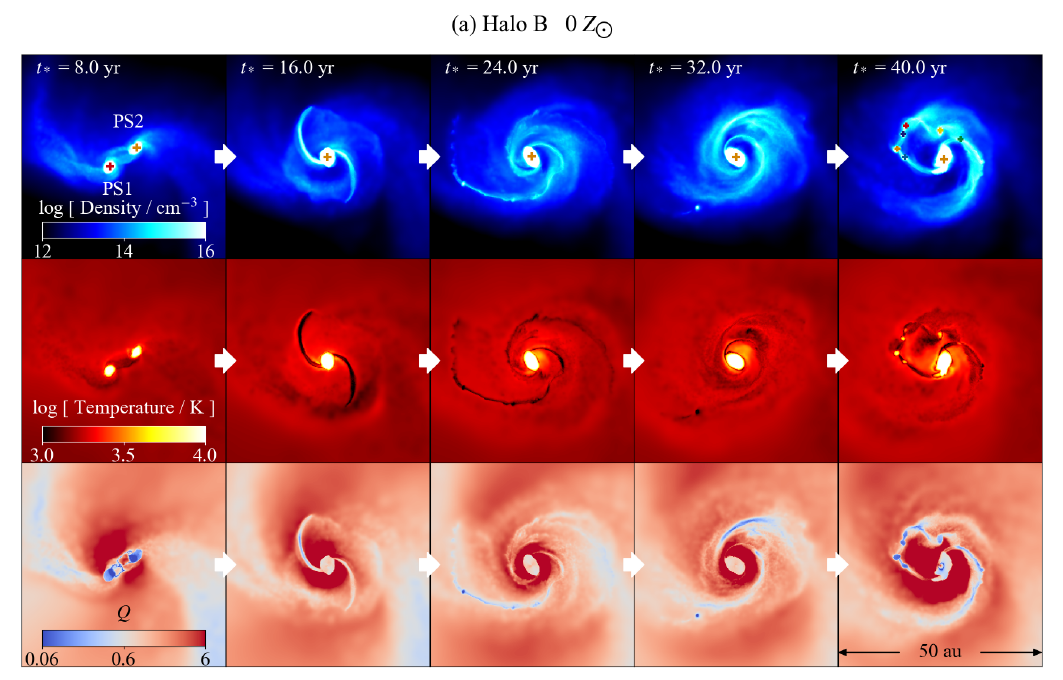}
 \includegraphics[width=18.0cm]{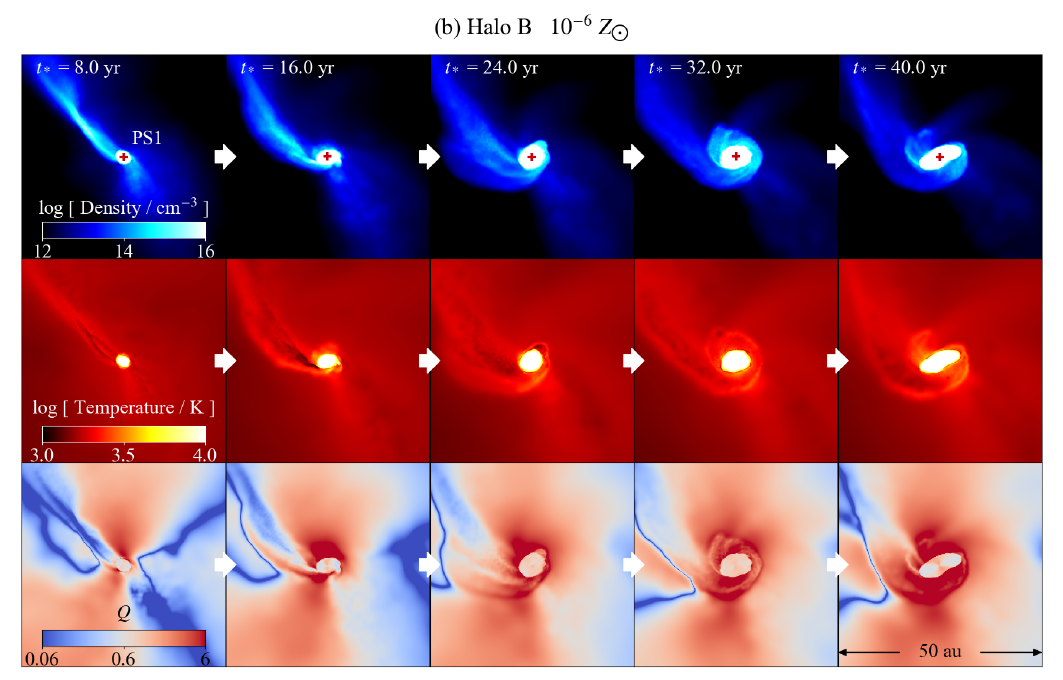}
\caption{
Projections of density (top panels), temperature (middle panels), 
and Toomre $Q$ parameter (bottom panels)
of the accretion discs for (a) {\tt HBZ0} and (b) {\tt HBZ-6} 
from $t_* = 8$ to 40 yr.
}
\label{fig:snapshots_acc_ini_H0611}
\end{figure*}

\subsection{Formation/destruction of protostars}

In this work, we find that multiple protostellar systems form in the accretion discs.
The protostars are created and destroyed through several processes as summarized in Section
\ref{sec:protostar_formation_destruction}.
In this section, we describe concrete cases of the formation/destruction processes.

\subsubsection{Formation of protostars}

Tables \ref{tab:PSs_HA} and \ref{tab:PSs_HB} show the formation time $t_{\rm form}$ and
the formation path of each protostar in our simulations for Halo A and B, respectively. 
We find that protostars form through the following three processes:
gravitational contraction (GC), the break-up of a protostar (BU), and 
the interaction of spiral arms (INT).

\vspace{0.5cm}
\noindent{{\bf (i) Gravitational contraction}}

Fig. \ref{fig:snapshots_acc_acc} shows the formation process of protostars through GC for {\tt HBZ-6}.
At $t_* = 8.3$, 9.5, and 11.5 yr, PS2, PS3, and PS4 form through the fragmentation of the filament
(Section \ref{sec:CF})
at distances 10.9, 12.0, and 11.2 au from PS1, respectively.
Linear analyses show that the most unstable wave number
of perturbations on an isothermal filament is $\lambda _{\max} = 2 \pi H$,
where $H = (2 \cs ^2 / \pi G \rho ) ^{1/2}$ is the scale height, or an effective radius
of the filament \citep{Nagasawa87, Inutsuka97}.
We can estimate the length-scale where fragmentation occurs to be $\lambda _{\max} = 16.2$ au 
for $\nH = 10^{14} \ \percc$ and $T = 1000$ K. 
The initial separation between the secondary protostars and PS1 is comparable to
this analytic estimate.
This is also consistent with the findings of our previous study \citepalias{Chiaki16}.

\vspace{0.5cm}
\noindent{{\bf (ii) Break-up of rapidly rotating protostars}}

Fig. \ref{fig:snapshots_acc_bu} shows the process of BU for {\tt HAZ-4}.
Since PS1 forms in the filament, it has a bar-like shape from the beginning.
After PS1 acquires sufficient angular momentum,
third or higher-order perturbations grow, and PS1 deforms into a
``dumb-bell-shape''.\footnote{More strictly, it is a Jacobi ellipsoid that
can have a ``dumb-bell-shape''
\citep{Eriguchi82}.
Protostars in our simulations are more oblate with aspect ratios 
$<0.25$, below the Jacobi sequence.
Further detailed modelling is needed to investigate a criterion for the break-up of such oblate objects
\citep{Tohline02}.}
The dumb-bell structure develops further and the protostar is eventually divided into two
separate ones.
In the classical theory of ``fission'', an object breaks up into an equal-mass binary \citep{Lyttleton53}.
In our run {\tt HAZ-4}, only 7\% of mass is taken apart from the main body,
which thus should be more appropriately characterized as ``mass-shedding'' suggested by \citet{Eriguchi82}. 

Protostars forming through BU show interesting accretion histories,
as shown in Figs \ref{fig:tm_H0522} and \ref{fig:tm_H0611}
(see 
the orange curve in Fig. \ref{fig:tm_H0522}b as a notable example).
Within $\sim 10$ yr after its formation,
the secondary protostar accretes gas rapidly
because it is still in a dense spiral arm.
Then the secondary protostar is ejected to a few tens au away from the center, 
where the gas density is as low as $10^{14} \ \percc$, and thus the accretion rate becomes low.
Since the initial ejection velocity is less than the escape velocity of the accretion
disc, the protostar falls back into the dense region and orbits around PS1.
The accretion rate increases again up to the value comparable to
that of PS1.

\vspace{0.5cm}
\noindent{{\bf (iii) Interaction of spiral arms}}

Fig. \ref{fig:snapshots_acc_int} shows the formation of a protostar
through the process of INT for {\tt HAZ-4}.
The spiral arms ARM1 and ARM2 collide with each other and produce a dense blob.
The blob eventually becomes self-gravitating, and a protostar PS3 forms at $t_* = 130$ yr.
To confirm the criterion for the fragmentation of spiral arms \citep[Eq. \ref{eq:Q};][]{Takahashi16, Inoue18}, 
we calculate the Toomre $Q$ parameter in ARM1 and ARM2.
The lower panels of Fig. \ref{fig:snapshots_acc_int} show that $Q$ is initially
around the critical value 0.6 on ARM1 and ARM2.
After the two arms collide, $\Sigma$ becomes larger by an order of magnitude.
Then, the minimum value of $Q$ becomes below 0.6, and the
dense region grows to a protostar PS3.
DF through INT have been studied by \citet{Inoue20}.
They perform a detailed analysis of the time evolution of spiral arms using
an analytic model and direct simulations to define the conditions for fragmentation.


\subsubsection{Destruction of protostars}

The destruction of protostars is an important process to regulate the number of
protostars.
Fig. \ref{fig:snapshots_acc_H0611_Z-9} shows
that, for {\tt HBZ0}, more than 20 protostars form within 50 yr
because of a high accretion rate 
($\simeq 0.04 \ \Msunyr$ at $t_* \simeq 40$ yr).
Most of the protostars merge with the massive protostars PS2, PS13 and PS20,
and only the three massive protostars survive at $t_* = 50$ yr.
We find the three characteristic destruction processes:
dissolution of protostars into ISM (DIS) mergers (MER) and TDEs.
Tables \ref{tab:PSs_HA} and \ref{tab:PSs_HB} show the destruction
time $t_{\rm dest}$ and path of each protostar. 
We show examples of MER and TDE in this section.


\vspace{0.5cm}
\noindent{{\bf (ii) Mergers with another member protostar}}

Fig. \ref{fig:snapshots_acc_mer} shows that a protostar PS3 merges with 
PS4 at $t_* = 92.3$ yr for {\tt HAZ-5}. 
PS3 approaches PS4 with an impact parameter $b = 0.13$ au.
It is smaller than the radius of PS4 $a_4 = 1.05$ au.
Through this near head-on collision, PS3 completely merges with PS4
(see Section \ref{sec:protostar_destruction}).

\vspace{0.5cm}
\noindent{{\bf (iii) Tidal disruption events}}

Some protostars are destroyed through TDEs.
Fig. \ref{fig:snapshots_acc_tde}a shows a TDE of PS4 with PS2 at $t_* = 90.0$ yr for {\tt HAZ-6}.
The impact parameter between them is $b = 1.59$ au.
It is larger than the radius of PS2 $a_2 = 0.48$ au,
but smaller than the tidal radius $r_{\rm tid} = 3.00$ au.
During the collision, PS4 is strongly deformed by the tidal force of PS2,
and dense arm-like structures form around PS2.
The whole elongated structure is eventually accreted onto PS4.
We also find a case where another protostar forms from the dense, deformed structure after TDEs.
Fig. \ref{fig:snapshots_acc_tde}b shows a TDE of PS4 with PS1 and subsequent formation of PS8
for {\tt HAZ-5}.
PS4 approaches PS1 at $t_* = 136$ yr with a large impact parameter $b = 3.82$ au compared to the protostellar radius $a_1 = 2.03$ au.
Just after the interaction, the specific angular momentum of PS1 becomes temporarily 
larger (corresponding to $j^2 = 0.028$) than the critical value 0.02 for break-up (see Fig. \ref{fig:tm_H0522}).
Although the impact parameter is larger than the tidal radius (2.30 au), TDE occurs in this case
(11th and 12th columns in Tables \ref{tab:PSs_HA} and \ref{tab:PSs_HB}).
Since the mass ratio between impacting and target protostars are comparable to the unity ($m_1/m_4 = 1.44$),
they deforms through each other's tidal force, and the cross-section increases.

\subsection{Case without fragmentation}
\label{sec:nofrag}
DF occurs for {\tt HBZ0} but not for {\tt HBZ-6},
although the thermal evolution is quite similar in these runs (Fig. \ref{fig:nT}).
Fig. \ref{fig:snapshots_acc_ini_H0611}
shows the density, temperature, and Toomre $Q$ parameter for {\tt HBZ0} and {\tt HBZ-6}, respectively.
For {\tt HBZ0}, a secondary protostar PS2 form at $t_* = 4.4$ yr
and collides with PS1 at at $10.2$ yr.
Since PS1 and PS2 have almost equal mass ($0.365 \ \Msun$ and $0.420 \ \Msun$, repsectively), the large amount of gas ($\sim 0.01 \ \Msun$) is stripped off into the ambient medium after the offset collision.
The gas is stretched up to $\sim 15$ au, and the 
temperature declines adiabatically in the elongated structure.
Then unstable spiral arms with $Q<0.6$ form (lower panels of Fig. \ref{fig:snapshots_acc_ini_H0611}a).

Contrastingly, for {\tt HBZ-6}, gas accretion onto the primary PS1 occurs
in an almost spherical manner 
(Fig. \ref{fig:rn_H0611}b).
Although multiple protostars form along the filament and collide with PS1,
these secondary protostars have small masses ($0.0782 \ \Msun$, $0.0568 \ \Msun$, and $0.0389 \ \Msun$)
relative to PS1 ($1.01 \ \Msun$ at $t_* = 14.4$ yr). 
The protostars collide with PS1 so quiescently that the material in the protostars
does not scatter.
Furthermore, since PS1 continues growing
with a high accretion rate ($\dot M_* \simeq 0.06 \ \Msunyr$),
the circumstellar medium is warm ($T\sim 10^4$--$10^5$ K), and 
is less susceptible for the gravitational instability (Fig. \ref{fig:snapshots_acc_ini_H0611}b).

\end{document}